\documentclass[twocolumn]{aastex631}

\usepackage[utf8]{inputenc}
\usepackage[T1]{fontenc}
\usepackage{CJK}

\newcommand\removed[1]{{}}

\def\Msun{\mathinner{{\rm M}_{\odot}}}

\def\ckpc{\mathinner{{\rm ckpc}}}
\def\pkpc{\mathinner{{\rm pkpc}}}
\def\kpc{\mathinner{{\rm kpc}}}

\def\cMpc{\mathinner{{\rm cMpc}}}

\def\Mpch{\mathinner{h^{-1}{\rm Mpc}}}
\def\cGpc{\mathinner{{\rm cGpc}}}

\def\Rvir{\mathinner{R_{200}}}

\def\Mstar{\mathinner{M_\ast}}
\def\asym{\mathinner{A}}
\def\sersic{\mathinner{n_{\rm S}}}
\def\sSFR{\mathinner{\rm sSFR}}
\def\Mtot{\mathinner{M_{\rm tot}^{\rm cl}}}
\def\Gyr{\mathinner{\rm Gyr}}
\def\Sersic{{S\'{e}rsic }}

\def\Om{\mathinner{\Omega_{\rm m}}}
\def\OL{\mathinner{\Omega_{\Lambda}}}
\def\Ob{\mathinner{\Omega_{\rm b}}}

\def\zf{\mathinner{z_{\rm f}}}

\begin{document}

\title{Emergence of the Galaxy Morphology-Star Formation Activity-Clustercentric Radius Relations in Galaxy Clusters}

\author[0000-0003-4923-8485]{Sungwook E. Hong}
\affiliation{Korea Astronomy and Space Science Institute,
776 Daedeok-daero, Yuseong-gu, Daejeon 34055,
Republic of Korea}
\affiliation{Astronomy Campus, University of Science and Technology,
776 Daedeok-daero, Yuseong-gu, Daejeon 34055,
Republic of Korea}

\author[0000-0001-9521-6397]{Changbom Park}
\affiliation{School of Physics, Korea Institute for Advanced Study,
85 Hoegiro, Dongdaemun-gu, Seoul 02455,
Republic of Korea}

\author[0009-0009-3772-3134]{Preetish K. Mishra}
\affiliation{School of Physics, Korea Institute for Advanced Study,
85 Hoegiro, Dongdaemun-gu, Seoul 02455,
Republic of Korea}

\author[0000-0002-4391-2275]{Juhan Kim}
\affiliation{Center for Advanced Computation, Korea Institute for Advanced Study, 85 Hoegiro, Dongdaemun-gu, Seoul 02455, Republic of Korea}

\author[0000-0003-4446-3130]{Brad K. Gibson}
\affiliation{Woodmansey Primary School, Hull Road, Woodmansey HU17 0TH, UK}

\author[0000-0003-4164-5414]{Yonghwi Kim}
\affiliation{Korea Institute of Science and Technology Information, 245 Daehak-ro, Yuseong-gu, Daejeon 34141, Republic of Korea}

\author{C. Gareth Few}
\affiliation{Woodmansey Primary School, Hull Road, Woodmansey HU17 0TH, UK}

\author[0000-0003-0695-6735]{Christophe Pichon}
\affiliation{Institut d'Astrophysique de Paris, CNRS, Sorbonne Université, UMR 7095, 98 bis bd Arago, 75014 Paris, France}
\affiliation{IPhT, DRF-INP, UMR 3680, CEA, L'orme des Merisiers, Bât 774, 91191 Gif-sur-Yvette, France}
\affiliation{School of Physics, Korea Institute for Advanced Study,
85 Hoegiro, Dongdaemun-gu, Seoul 02455,
Republic of Korea}

\author[0000-0001-5135-1693]{Jihye Shin}
\affiliation{Korea Astronomy and Space Science Institute,
776 Daedeok-daero, Yuseong-gu, Daejeon 34055,
Republic of Korea}
\affiliation{Astronomy Campus, University of Science and Technology,
776 Daedeok-daero, Yuseong-gu, Daejeon 34055,
Republic of Korea}

\author[0000-0002-6810-1778]{Jaehyun Lee}
\affiliation{Korea Astronomy and Space Science Institute,
776 Daedeok-daero, Yuseong-gu, Daejeon 34055,
Republic of Korea}



\begin{abstract}
We investigate when and how the relations of galaxy morphology and star forming activity with clustercentric radius become evident in galaxy clusters. We identify 162 galaxy clusters with total mass $\Mtot > 5 \times 10^{13} \Msun$ at $z = 0.625$ in the {\it Horizon Run 5} (HR5) cosmological hydrodynamical simulation and study how the properties of the galaxies with stellar mass $\Mstar > 5 \times 10^9 \Msun$ near the cluster main progenitors have evolved in the past. Galaxies are classified into disk, spheroid, and irregular morphological types according to the asymmetry and \Sersic index of their stellar mass distribution. We also classify galaxies into active and passive ones depending on their specific star-formation rate. We find that the morphology-clustercentric radius relation (MRR) emerges at $z \simeq 1.8$ as the fraction of spheroidal types exceeds 50\% in the central region ($d \lesssim 0.1 \Rvir$). Galaxies outside the central region remain disk-dominated. Numerous encounters between galaxies in the central region seem to be responsible for the morphology transformation from disks to spheroids. We also find that the star formation activity-clustercentric radius relation emerges at an epoch different from that of MRR. At $z\simeq0.8$, passive galaxies start to dominate the intermediate radius region ($0.1\lesssim d/\Rvir \lesssim0.3$) and this ``quenching region'' grows inward and outward thereafter. The region dominated by early-type galaxies (spheroids and passive disks) first appears at the central region at $z\simeq 1.8$, expands rapidly to larger radii as the population of passive disks grows in the intermediate radii, and clusters are dominated by early types after $z\simeq 0.8$.

\end{abstract}



\section{Introduction} \label{sec:intro}

Galaxies in the nearby universe exhibit tremendous diversity in their structure and star formation properties. The necessity for a systematic study of galaxy populations has led astronomers to classify galaxies into various morphological classes \citep{Hubble1926, deVaucouleurs1959, Sandage1975}. The low-redshift galaxy population can be divided into spheroidal ellipticals, disky lenticulars (S0s), spirals, and irregulars. The elliptical and S0 galaxies are often grouped together and called early-type galaxies (ETGs), while spirals and irregulars are combined into late-type galaxies (LTGs) \citep{park2005}. Despite our extensive knowledge of the detailed structural and star formation properties of these different morphological classes, we do not fully understand the origin of their diversity. Recent studies of galaxy morphology from observations and simulations have shown that the vast majority of the low-mass $(M_* < 10^{10} M_{\odot})$ galaxies at high redshifts ($z \geq 2$) are dominated by disks, with only minor contributions from spheroids and irregulars 
\citep{park2022, Lee2024}. This result, when juxtaposed with the fact that the majority of massive low-redshift galaxies are early types \citep{Kelvin2014}, points to the morphological transformation of high-redshift galaxies with cosmic time. Understanding the underlying factors that govern this morphological evolution of galaxies is an active area of research.

One crucial factor that has been argued to impact morphology is the surrounding environment of galaxies. Observations have revealed that early-type galaxies are more abundant in galaxy clusters than in the field environment. \citet{dressler1980} found that the fraction of ETGs increased smoothly as one moved from low to high locally dense regions or radially toward the cluster center. These connections are known as the morphology-density relation (MDR) and morphology-radius relation (MRR), respectively. The MDR was later generalized to group and field environments \citep{goto2003} and was observed to be in place up to redshift $z\sim 1$ \citep{Smith2005, hwang2009}. The validity of MDR in galaxy clusters and also in the field environment, however, has been questioned in several works. Research on low-to-intermediate redshift galaxy clusters has shown that the clustercentric distance, often normalized by a characteristic cluster radius, is a more important factor related to morphological fractions than the local environmental density \citep{whitmore1993, treu2003, park2009, Fasano2015}. 

The idea that MDR is a fundamental relation extending from field to group to cluster environments has also been scrutinized. Several studies \citep[e.g.,][]{park2007, park2008, Park&Choi2009} investigated the dependence of various galaxy properties, including morphology, on small- and large-scale environments at fixed galaxy absolute magnitude or mass. They found that the environmental effects on the galaxy morphology are best described by two dominant factors: the distance to the nearest neighbor ($d_{\rm nei}$) and its morphology, rather than the local galaxy number density smoothed over a $3$ to $8\Mpch$ scale \citep{park2007}. When the nearest neighbor environment is fixed, galaxy morphology does not vary with the local density. Moving into galaxy clusters, they found that, at fixed mass, the morphology of a cluster galaxy is mainly determined by the clustercentric radius. Only when $d_{\rm nei}$ is very small ($d_{\rm nei}/R_{\rm vir,nei} \lesssim 0.1$, where $R_{\rm vir,nei}$ is the neighbor's virial radius), galaxy morphology is determined by $d_{\rm nei}$ and its neighbor's morphology \citep{park2009}. According to this study, the MDR is an apparent phenomenon arising from the statistical correlation of the local density with galaxy mass and the nearest neighbor environment.

Despite its importance over MDR, the study of MRR and its redshift evolution has received less attention. Observations show that MRR is well-established at low redshift, slightly stronger in relaxed clusters than in irregular/clumpy clusters \citep{dressler1997, Fasano2015}. The major population of ETGs also seems to be in place at redshift $z \sim 1$ and shows weak evolution over cosmic time with most of the evolution happening only in the outskirts \citep{Huertas-Company2009}. Due to the limited high-redshift observations of MRR and the lack of related attention in simulations, our understanding of the origin and evolution of MRR remains quite limited.

It must be noted that a complete understanding of MRR in clusters requires understanding not only the evolution of galaxy structure but also the star formation activity therein. For example, the lenticulars, the main constituent of the cluster population, are structurally similar to spirals \citep[e.g.,][]{Barway2009, Laurikainen2010, Mishra2019}, but are labelled as ETGs because of their red color or passive star-formation activity. Therefore, the radial distribution of the specific star formation rate (sSFR) of cluster galaxies should be explored together with the MRR. The study of the star formation activity-clustercentric radius relation (ARR) of galaxies is interesting in its own right. For low redshift clusters, it is known that cluster members have lower SFR compared to field galaxies. However, the epoch of cosmic time at which such a difference in star formation properties between cluster and field galaxies arises is a topic of active debate. Several studies find the sSFR of cluster galaxies to be largely consistent with that of field galaxies \citep[e.g.,][]{Brodwin2013, Alberts2016, Trudeau2024} at $z\gtrsim1.5$, while a few others \citep[e.g.,][]{Cooke2019, Strazzullo2019} claim significant differences in sSFR between the two populations at this redshift.

From these aforementioned findings, it is crucial to factor out star formation activities in the morphological classification to fully understand the MRR of clusters. For the purpose of this study, we adopt a convention in which we define galaxy morphology purely based on galaxy structure and classify galaxies into spheroids, disks, and irregulars. However, we also use a combination of galaxy structural parameters and star formation rate to classify galaxies into early or late types.

In this work, we explore the emergence of the galaxy morphology-star formation activity-clustercentric radius relation using the \textit{Horizon Run 5} \citep{lee2021} cosmological hydrodynamical simulation. Because of its large volume and high resolution, the \textit{Horizon Run 5} simulation has been successful in accurately predicting high-redshift galaxy morphology \citep{park2022}, which has been verified using the data from \textit{James Webb Space Telescope} (JWST) observations \citep{Lee2024}. This makes \textit{Horizon Run 5} an ideal simulation to study the morphological evolution of galaxies in and around clusters as a function of cosmic time.

The structure of this paper is as follows. In Section~\ref{sec:data}, we describe the mock galaxy samples from the {\it Horizon Run 5} cosmological hydrodynamic simulation and how we assign the morphology and star formation activity types to our galaxy samples. We then show how various physical parameters of cluster galaxies evolve in various environments in Section~\ref{sec:results}. Finally, we summarize our results in Section~\ref{sec:conclusion}.

\section{Data}\label{sec:data}

In this section, we briefly describe the \textit{Horizon Run 5} simulation, the identification of halos and galaxies, their merger trees, and morphology and star formation activity type classifications. For details, readers may refer to \citet{lee2021, park2022, kim2023} and the Appendix.

\subsection{Horizon Run 5 Simulation}\label{sec:data_HR5}

\textit{Horizon Run 5} \citep[HR5 hereafter;][]{lee2021} is a state-of-the-art cosmological hydrodynamic simulation performed in an extremely large volume of $1.15 \cGpc^3$, containing a spatial resolution of up to $\sim 1\kpc$ for the zoom-in region of $(1049 \cMpc) \times (119 \cMpc) \times (127 \cMpc) = 1.58 \times 10^{-2} \cGpc^3$. HR5 adopted a flat $\Lambda$CDM cosmological model best fit to the \textit{Planck} 2015 data \citep{planck2016}: $(\Om, \OL, \Ob, \sigma_8, h) = (0.3, 0.7, 0.047, 0.816, 0.684)$.

HR5 was run using a modified version of the adaptive mesh refinement code \texttt{RAMSES} \citep{teyssier2002,lee2021}, which outputs the evolution of dark matter (DM), star, massive black hole (MBH) particles, and gas cells. New highest levels of refinement have been added at $z = 79$, $39$, $19$, $9$, $4$, and $1.5$ to maintain the highest spatial proper space resolution to $\Delta x \sim 1\pkpc$ until the simulation reached its final redshift at $\zf = 0.625$. For details, readers may refer to Appendix~\ref{sec:app_data_HR5}.

\subsection{Cluster Galaxy \& Progenitor Identification}\label{sec:data_galaxy}

At each HR5 snapshot, halos were identified by applying the Friends-of-Friends (FoF) method to DM/star/MBH particles and gas cells. Then, galaxies were identified by applying a substructure-finding algorithm \texttt{PGalF} \citep{lee2021, kim2023} to the member star particles of each halo. Both halos and galaxies at different time steps were connected by merger trees, where progenitors of a given halo and galaxy were defined by using the most bound particles \citep[MBPs;][]{hong2016} and an updated version of \texttt{ySAMtm} code \citep{park2022, lee2014, jung2014}, respectively. In this paper, we only use the most massive progenitors in each snapshot of given halos and galaxies and call them progenitors. For details, readers may refer to Appendices~\ref{sec:app_data_galaxy} and \ref{sec:app_data_tree}.

We further define ``clusters'' and ``cluster galaxies,'' subsets of HR5 halos and galaxies at $z = \zf$, as follows. First, clusters are defined as halos in the high-resolution hydrodynamic region with a sufficiently large total mass 
\begin{equation}
\Mtot \equiv M_{\rm DM}^{\rm cl} + M_\ast^{\rm cl} + M_{\rm MBH}^{\rm cl} + M_{\rm gas}^{\rm cl} > 5 \times 10^{13} \Msun
\label{eq:mtot_def}
\end{equation}
at $z = \zf$, with individual mass components being the mass of DM, stars, MBH, and gas, respectively. Then, cluster galaxies are defined as clusters' member galaxies with sufficiently large stellar mass $\Mstar > 5 \times 10^9 \Msun$ at $z = \zf$ to make sure that our morphology estimation is reliable.

To study the impact of cluster environments on the cluster member galaxies, we need to trace back the history of both the cluster and its member galaxies. We define the progenitors of clusters as the most massive progenitor halo along the merger tree. The progenitors of low-redshift cluster galaxies were identified as the most massive galaxy progenitors using merger trees. We note that not all cluster galaxies originated within the cluster progenitors. This is especially true for low-mass cluster galaxies and galaxies residing on the outskirts of clusters, many of whose progenitors originated far away from the cluster progenitors. However, these galaxies are crucial for understanding how the properties of galaxies evolve once they become members of clusters.

Next, to study the evolution of galaxies as a function of distance to the cluster center, we first define centers of clusters and cluster-progenitors as the local density peak positions of the total mass from their most massive substructure. We found that, rather than simply using the distance ($d$) from the cluster center, a scaled distance normalized by the virial radius of the cluster ($\Rvir$) is more relevant in studying the environmental effects in the cluster. We measured $\Rvir$ by iteratively changing the radius of a spherical kernel positioned at the cluster center until we obtain the average density to be 200 times the critical density ($\rho_{\rm c}$). However, rather than using different values of $\Rvir$ along the history of the target cluster, we fixed it to the value measured at $z=\zf$ on the comoving scale. This is because, as mentioned before, we have used the normalized clustercentric distance, which was devised to scale the results among different-mass cluster samples. Therefore, we want to fix the comoving radius for each cluster to study the connection between galaxy morphology and sSFR along the history line.

From now on, we drop the term ``progenitor'' and, therefore, whenever we say a ``cluster'' and ``cluster galaxies'' at $z>\zf$, they mean the progenitor of a cluster and the progenitors of cluster galaxies, respectively. In addition to the cluster galaxies, we measure morphology and sSFR of the field galaxies outside clusters as they are important to contrast the relationships between galaxy properties and environment. Whenever we use such galaxies in our analysis, we will explicitly mention them hereafter.

\subsection{Galaxy Classifications}\label{sec:data_morph_type}

We use the same method described in \citet{park2022} to measure the ``morphology'' of HR5 cluster galaxies. Here, the galaxy morphology is defined by using two parameters: first, the ``asymmetry'' parameter $A$ \citep{conselice2000, park2022} is defined as a relative difference of stellar mass density between points reflected with respect to the galaxy center (see Appendix~\ref{sec:app_data_morph} for mathematical definition). Following \citet{park2022}, we classify all galaxies into ``symmetric'' ($\asym < 0.4$) and ``asymmetric'' ($\asym > 0.4$). For the symmetric galaxies, we calculate the \Sersic index $\sersic$ of the radial stellar mass density profile in the face-on view of each galaxy \citep{sersic1963} to distinguish them into ``disks'' ($\sersic < 1.5$) and ``spheroids'' ($\sersic > 1.5$). For this calculation, we exclude the central region of galaxies within the diameter of $1.6 \pkpc$, as this region is not well resolved in HR5, and also to avoid the possible central bulge region. Some interacting galaxies show a non-monotonic density profile that makes the \Sersic profile fitting fail and are classified as ``irregulars'' together with the asymmetric ones. For details, readers may refer to Appendix~\ref{sec:app_data_morph}. The morphology measure, as noted in the reference, differs from those adopted in observations that use the projected light distribution rather than the projected stellar-mass distribution. However, our morphology provides the physical properties of galaxies, which could serve as a proxy for understanding the origin and evolution of the galaxy morphology \citep{trayford2019}.

Together with the morphological types, we also reclassify the HR5 cluster galaxies into ``early'' and ``late'' types for a direct comparison with low-$z$ observational results of galaxy types in various environments, including clusters \citep[e.g.,][]{dressler1980, park2007, hwang2009}. It is worthwhile to note that, although our morphology is defined solely by the spatial distribution of star particles, observational classifications of early- and late-type galaxies are influenced by various physical properties, primarily their star formation activity that is cumulatively reflected in their colors or spectra. For example, lenticular (S0) galaxies are classified into disks in our study because they host a well-developed disk component ($\sersic \simeq 1$). However, in terms of color and spectral features, they are classified as early-type-like ellipticals due to their old stellar populations.

\begin{deluxetable}{ll}
\caption{Summary of three galaxy classification schemes used in this paper. $\asym$, $\sersic$, and $\sSFR$ correspond to asymmetry parameter, \Sersic index, and specific star formation rate, respectively. See the texts for details.}
\label{tab:morphology_type_summary}
\tablehead{\colhead{Name} & \colhead{Description}}
\startdata
Disks & $\asym < 0.4, \sersic < 1.5$ \\
Spheroids & $\asym < 0.4, \sersic > 1.5$ \\
Irregulars & $\asym > 0.4$ 
or null-$\sersic$ \\
\hline
Active Galaxies & $\sSFR > 0.05 \Gyr^{-1}$ \\
Passive Galaxies & $\sSFR < 0.05 \Gyr^{-1}$ \\
\hline
Early-type Galaxies & Spheroids or Passive Disks \\
Late-type Galaxies & Irregulars or Active Disks \\
\enddata
\end{deluxetable}

In this paper, we adopt a classification similar to \citet{park2005} and \citet{park2022} to define ETGs and LTGs. First, depending on their specific star formation rate ($\sSFR \equiv {\rm SFR} / \Mstar$), we divide our galaxy sample into ``active'' ($\sSFR > 0.05 \Gyr^{-1}$) and ``passive'' ($\sSFR < 0.05 \Gyr^{-1}$) galaxies (see Appendix~\ref{sec:app_data_active_passive} for details). We then combine spheroids ($\asym < 0.4$, $\sersic > 1.5$) and passive disks, a proxy of S0 galaxies ($\asym < 0.4$, $\sersic < 1.5$, $\sSFR < 0.05 \Gyr^{-1}$), to call them ETGs. On the other hand, active disks ($\asym < 0.4$, $\sersic < 1.5$, $\sSFR > 0.05 \Gyr^{-1}$) and irregulars ($\asym > 0.4$ or null value of $\sersic$) are called LTGs. Table~\ref{tab:morphology_type_summary} summarizes our galaxy morphology, star-forming activity, and early-/late-type classifications. Another scheme dividing galaxies into fast and slow rotators can be added depending on their $v/\sigma$, where $v$ is the systematic rotation velocity and $\sigma$ is the central velocity dispersion. It should be noted that the tight correlations between these types observed at low redshifts are the result of the evolution of galaxies throughout the history of the universe and are not necessarily strong at high redshifts. Therefore, one should not use the parameters of one classification scheme as proxies for another classification scheme when high-redshift galaxies are studied.

\section{Results}\label{sec:results}

\subsection{Global Evolution}\label{sec:results_global}

\begin{figure*}[tb]
    \centering
    \includegraphics[width=0.32\textwidth]{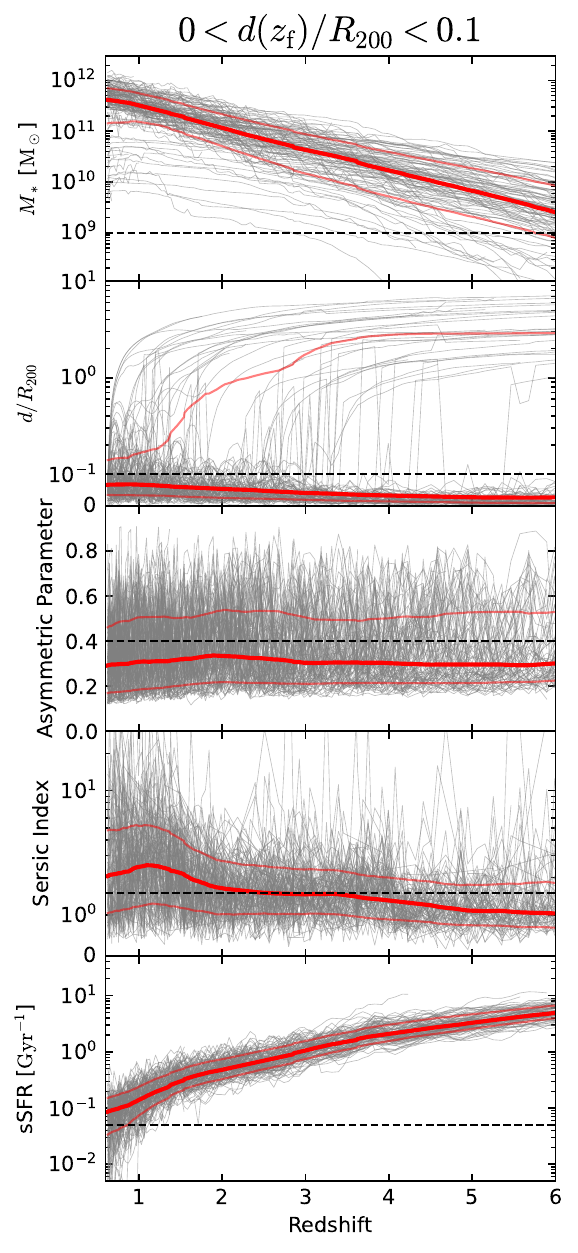}
    \includegraphics[width=0.32\textwidth]{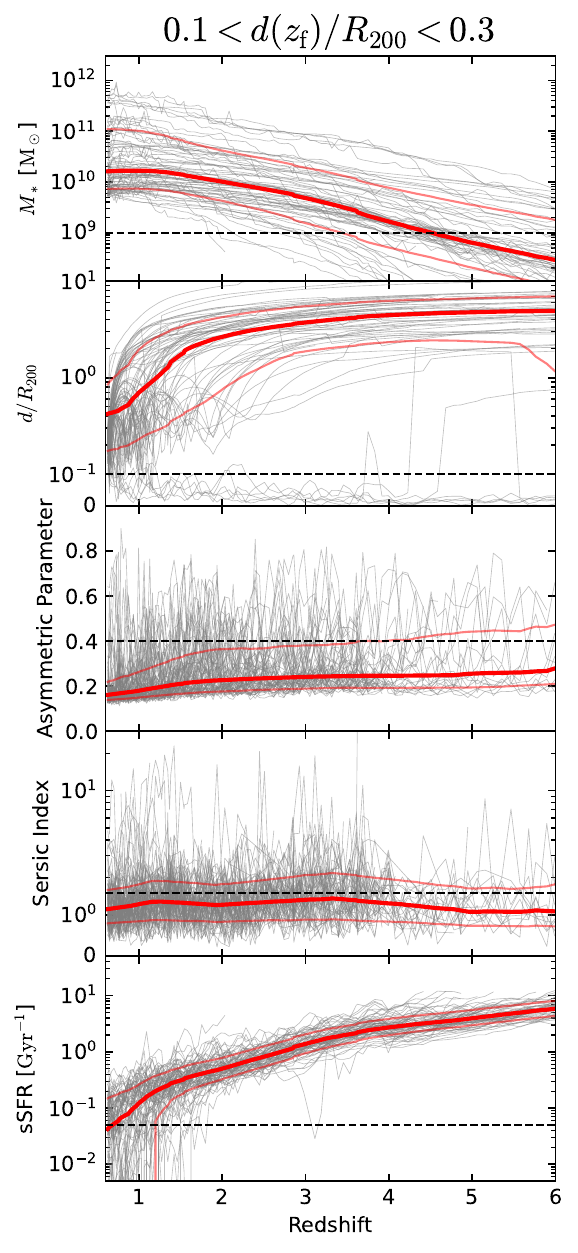}
    \includegraphics[width=0.32\textwidth]{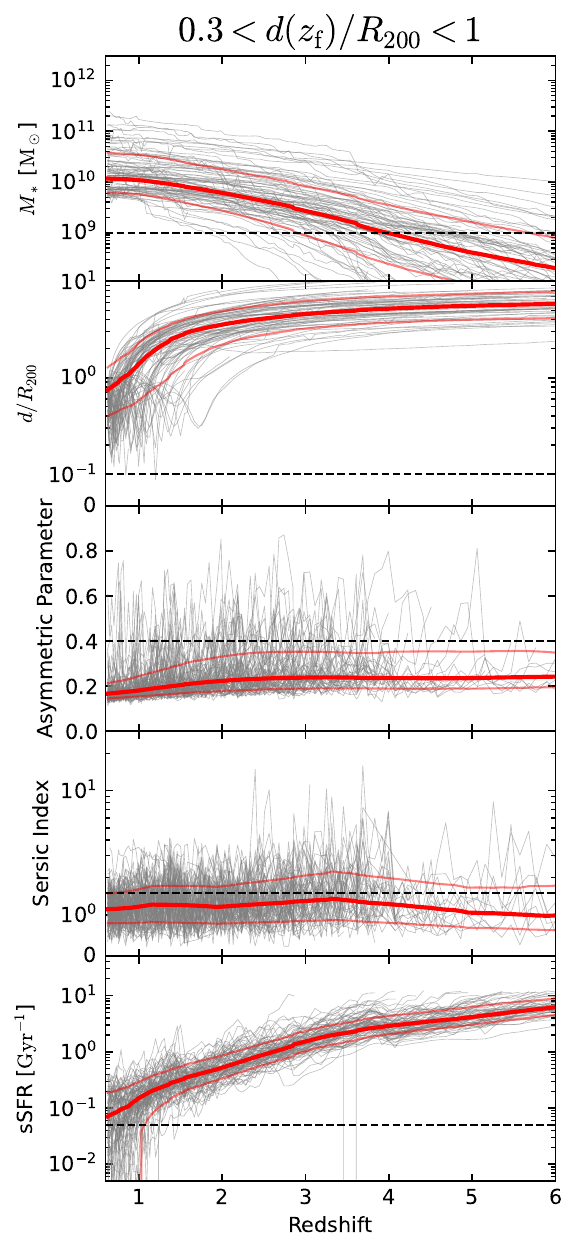}
    \vspace{-10pt}
    \caption{Evolution of various physical parameters of HR5 cluster galaxies from $z = 6$ to $0.625$. From top to bottom panels: stellar mass ($\Mstar$), normalized clustercentric radius ($d/\Rvir$), asymmetric parameter ($\asym$), \Sersic index ($\sersic$), and specific star formation rate (sSFR). Note that the hosting clusters and the comoving values of their virial radii $\Rvir$ are defined as those at $z = 0.625$.
    From left to right panels: $d(\zf) / \Rvir < 0.1$, $0.1 < d(\zf) / \Rvir < 0.3$, and $0.3 < d(\zf) / \Rvir < 1$. Thick and thin red lines show the median and 68\% percentiles, respectively. Thin gray lines show the histories of 100 selected individual galaxies in each normalized clustercentric radius bin. Dashes from top to bottom panels: $\Mstar = 10^9 \Msun$, $d = 0.1 \Rvir$ (central region), $\asym = 0.4$ (regular/irregular threshold), $\sersic = 1.5$ (disk/spheroid threshold), and $\sSFR = 0.05 \Gyr^{-1}$ (active/passive threshold).}
    \label{fig:1d}
\end{figure*}

Figure~\ref{fig:1d} shows the global evolution of stellar mass, the normalized clustercentric radius ($d/\Rvir$), the asymmetry parameter, the \Sersic index, and the sSFR of HR5 cluster galaxies from $z = 6$ to $0.625$. $\Rvir$ is fixed to the value at $z=0.625$ in the comoving scale at all redshifts for each cluster. We separated our cluster galaxies into three bins according to their normalized clustercentric radius at the final snapshot --- $d(\zf)/\Rvir < 0.1$ (central), $0.1 < d(\zf)/\Rvir < 0.3$ (inner shell), and $0.3 < d(\zf)/\Rvir < 1$ (outer shell). Interestingly, the overall trends of the global evolution of the structural properties of the ``outskirts''
galaxies in the inner and outer shells are similar, while those of the central galaxies are quite different from the others. We highlight the differences in the following paragraphs.

First, central galaxies always have relatively large stellar mass from cosmic morning ($z>4$) to the last snapshot ($\Mstar \simeq 3 \times 10^9 \Msun$ at $z = 6$; $\Mstar \simeq 3 \times 10^{11} \Msun$ at $z = \zf$), while most of the outskirts galaxies always have been low-mass galaxies throughout the history ($\Mstar \simeq 3 \times 10^8 \Msun$ at $z = 6$; $\Mstar \simeq 10^{10} \Msun$ at $z = \zf$). Not only the overall stellar mass level but also the increment of the stellar mass is different between central and outskirts galaxies. For example, the median stellar mass of central galaxies increases about $10^2$ times from $z = 6$ to $\zf$, while those of outskirts galaxies increase only $\sim 30$ times during the same period. Such difference implies that the primary mechanisms of stellar mass growth and their operation epochs are quite different between central and outskirts galaxies \citep{park2022}.

Second, most of the outskirts galaxies originated far outside the clusters ($d \gtrsim 3 \Rvir$) and entered the virialized range of their host clusters after $z \sim 1$. In contrast, most of the central galaxies have always been close to their cluster centers ($d \ll 0.1 \Rvir$) throughout history, while only a small fraction of them have been around cluster outskirts, possibly due to the major merger events of galaxies and halos (see Appendix~\ref{sec:app_flow_precise}, for example).

Third, because of the higher merger rate, central galaxies have a median value of $\textrm{med}(\asym)\simeq 0.3$ significantly higher than the outskirt case ($\textrm{med}(\asym)\simeq 0.2$) while the both of them always have the median asymmetricity lower than $\textrm{med}(\asym)=0.4$. About one-third of central galaxies (and their progenitors) are asymmetric ($\asym > 0.4$) throughout merger history, while the fraction is always $< 10\%$ for outskirts galaxies. The distribution of $\asym$ of the outskirts galaxies also becomes increasingly narrower with a decrease in redshift (especially at $z \lesssim 2$), which shows that more outskirts galaxies have acquired a regular morphology with time. \citet{park2022} showed that high-$\asym$ is closely related to either major merger events or close encounters of galaxies in the cosmic morning, and we expect similar origins for the relatively high fraction of asymmetric galaxies around cluster centers.

The inner- and outer-shell galaxies seem to share the same evolution of the \Sersic index at all redshifts. On the other hand, the median value of the \Sersic index shows differences between them.  Centrals start to have a higher $\textrm{med}(\sersic)$ after $z\sim 3.5$ with a difference steadily increasing until $z\sim 1.8$, after which it significantly diverges. As a result, most of the central galaxies have $\sersic >1.5$ becoming spheroidal at the final redshift. To the contrary, the majority of the outskirts galaxies remain disky at all redshifts.

We conclude that most central galaxies are massive ($\Mstar \gtrsim 10^{11} \Msun$) and spheroidal ($\sersic > 1.5$), mainly the brightest cluster galaxies (BCGs) throughout history (with $d < 0.1 \Rvir$), suffered numerous major mergers and close encounters ($\sim 1/3$ having $\asym > 0.4$). On the other hand, most of the outskirts galaxies are low-mass ($\Mstar \lesssim 10^{11} \Msun$) and disky ($\sersic < 1.5$) having originated from outside of the clusters.

We now describe the global evolutionary trend of the sSFR of cluster galaxies. The global sSFR of all galaxies has steadily declined from high to low redshifts. The evolutionary trend of sSFR is nearly the same for galaxies in all clustercentric radius bins until $z \sim 1.5$. After this epoch, a substantial fraction of outskirts galaxies become passive, while a major fraction of central galaxies remain star-forming until the final snapshot. Among the outskirts galaxies, the tendency to become passive is stronger in inner-shell galaxies, with a nonnegligible fraction of them already passive by $z \sim 1$. The outer shell galaxies show a relatively slow decline in star formation activity, with their median sSFR crossing the threshold for passive evolution only around $z \sim 0.8$.

\begin{figure}[tb]
    \centering
    \includegraphics[width=0.48\textwidth]{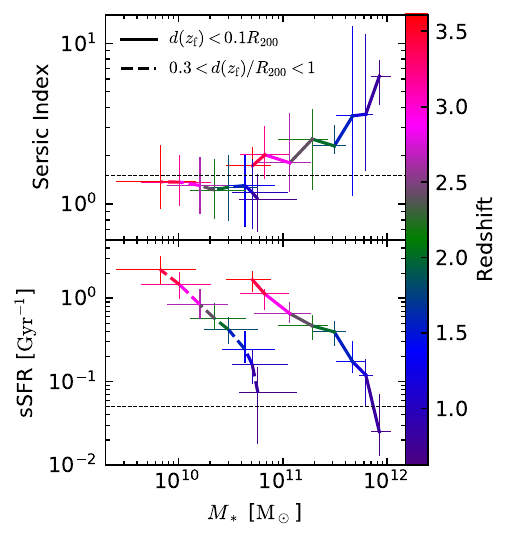}
    \vspace{-20pt}
    \caption{Time evolution of \Sersic index (top) and sSFR (bottom) with stellar mass for top-20\% massive HR5 central galaxies ($d(\zf) < 0.1 \Rvir$; solid) and outskirts galaxies ($0.3 < d(\zf) / \Rvir < 1$; dashed) from $z = 3.6$ (red) to $0.625$ (purple).
    Thick lines and thin crossbars correspond to the median and 68\% percentiles at the given time, respectively.}
    \label{fig:mstar_sersic_ssfr_history}
\end{figure}

Note that in the cosmic morning and early noon ($z\gtrsim3$), there is not much difference in the asymmetry parameter, \Sersic index, and sSFR between central and outskirts galaxies, while there exists distinguishable difference in the stellar mass distribution. This indicates that stellar mass may not be the sole driver for the evolution of galaxy morphology. Similar results can be seen in Figure~\ref{fig:mstar_sersic_ssfr_history}, which shows the time evolution of $\Mstar$-$\sersic$ and $\Mstar$-$\sSFR$ relations for top-20\% massive central and outskirts galaxies. If the galaxy stellar mass is the main driver of morphology evolution \citep[e.g.,][]{brough2017}, central and outskirts galaxies should have a similar morphology at the same stellar-mass scale. To the contrary, we found that both $\sersic$ and sSFR of central and outskirts galaxies are significantly different, even at the same-mass scale (see $\sersic$ and sSFR at $\log \left(\Mstar/\Msun\right) \simeq 10.8$ in Figure~\ref{fig:mstar_sersic_ssfr_history}). In light of this argument, the structural differences seen in central versus outskirts galaxies imply that clustercentric radius or related environmental parameters also affect the evolution of galaxy morphology \citep[see, e.g.,][]{Park&Choi2009, park2009, peng2010, bluck2019}.

\begin{figure*}[tb]
    \centering
    \includegraphics[width=0.45\textwidth]{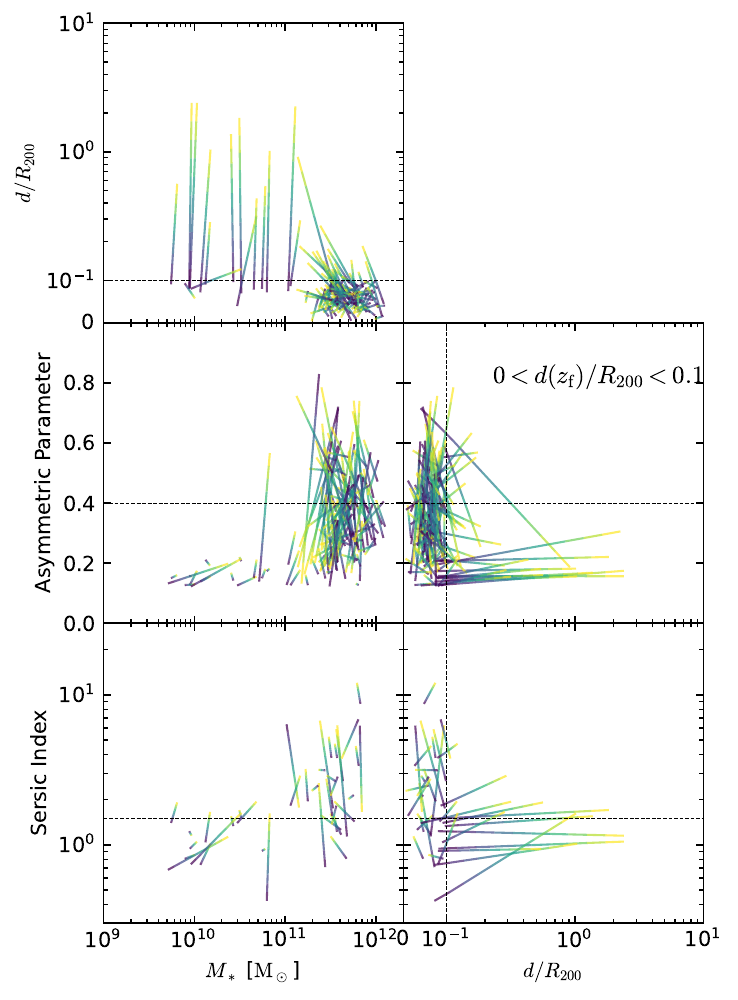}
    \hspace{10pt}
    \includegraphics[width=0.45\textwidth]{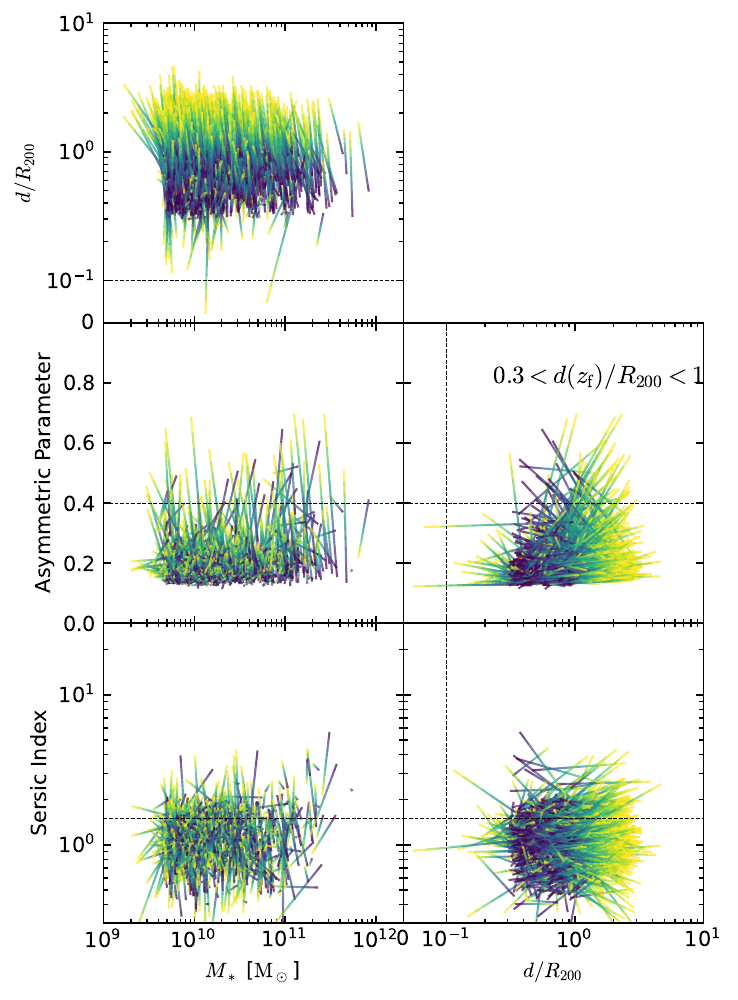}
    \vspace{-10pt}
    \caption{Evolution of various parameter-parameter relations between $z = 1$ (yellow) and $0.625$ (dark blue) for HR5 central (left) and outskirts (right) galaxies. From top-left to counter-clockwise direction in each panel: $\Mstar$-$(d/\Rvir)$, $\Mstar$-$\asym$, $\Mstar$-$\sersic$, $(d/\Rvir)$-$\sersic$, and $(d/\Rvir)$-$\asym$. Note that the number of lines associated with $\sersic$ is less than others because $\sersic$ can be calculated only for galaxies having $\asym < 0.4$.}
    \label{fig:flow}
\end{figure*}

The previous discussion on the global evolution of the physical properties of cluster galaxies makes it clear that, on average, central and outskirts galaxies have developed different morphologies since around $z \sim 3.5$. Now, we want to explain how the evolutionary history of individual galaxies may shape the evolutionary path of morphology at $0.625\leq z\leq 1$. Figure~\ref{fig:flow} shows the changes of galaxy parameters for central (left panel) and outer-shell (right panel) galaxies between $z=1$ (yellow) and $\zf$ (dark blue). Among our HR5 cluster galaxies, most of the very massive ones ($\Mstar \gtrsim 10^{11} \Msun$ at $z = \zf$) have been concentrated at $d \lesssim 0.2 \Rvir$ during the time while only a few of them came from outskirts (see $\Mstar$-$(d/\Rvir)$ panels). Such massive central galaxies show a variety of asymmetry parameter values ($\asym \approx 0.1-0.8$), which suggests that frequent major merger events may have occurred around cluster centers \citep[see $\Mstar$-$\asym$ and $(d/\Rvir)$-$\asym$ panels;][]{park2022}. Also, most of such massive central galaxies have a high value of \Sersic index ($\sersic \gtrsim 2$), which satisfies our expectations that BCGs suffering numerous major merger events would have spheroidal morphology \citep[see $\Mstar$-$\sersic$ and $(d/\Rvir)$-$\sersic$ panels;][]{liu2015,kluge2023}. Note that while each galaxy may have either increment or decrement of $\sersic$, an overall behavior between $\Mstar$ (or time) and $\sersic$ shows a positive relation. That is, although each major merger event may randomly change the morphology of a given massive central galaxy, multiple major merger events would make such galaxies more spheroidal in stochastic ways \citep[e.g.][]{Martin2018}. Such a stochastic change toward spheroidal shape seems stronger for massive galaxies \citep[$\Mstar \gtrsim 10^{11} \Msun$;][]{rodriguez2017}. And restoration of morphology from spheroid back to disk by the angular momentum supplied by the systematic inflow of gas and small galaxies under the large-scale tidal torque \citep{park2022} is difficult to happen for massive galaxies that are located at the central nonlinear region of the cluster dark matter halo.

On the other hand, relatively low-massive ones ($\Mstar \lesssim 10^{11} \Msun$ at $z = \zf$) show a variety of clustercentric radii at $z = 1$ regardless of their final positions, with a large fraction of them coming from outside of the virial radii (see $\Mstar$-$(d/\Rvir)$ panels). Also, regardless of their initial values at $z = 1$, both the asymmetry parameter and the \Sersic index end up around their typical range of $\asym = 0.1-0.3$ and $\sersic = 0.5-2$. The mild decrease in the median value of the \Sersic index of central galaxies as observed in Figure~\ref{fig:1d} at $z\lesssim1$ is likely driven by such low-mass galaxies originating beyond cluster radius that have become a part of central galaxy population at later epochs. We refer the readers to Appendix~\ref{sec:app_flow_precise} for the detailed evolution of parameter-parameter relations for selected individual galaxies.

\subsection{Evolution of MRR and ARR}\label{sec:result_radius_morph_type}

In the previous subsection, we have seen that massive, central galaxies ($\Mstar \gtrsim 10^{11} \Msun$; $d < 0.1 \Rvir$ at $z = \zf$) have changed their morphology from disks to spheroids in a stochastic manner by numerous events of major merger and close encounters occurred near the cluster centers. On the other hand, low-mass outer shell galaxies ($\Mstar \lesssim 10^{11} \Msun$; $d > 0.3 \Rvir$ at $z = \zf$) have not had a sufficient amount of such events around cluster outskirts and, therefore, tend to remain disks. In this subsection, we will study how such different merger/close encounter histories of cluster galaxies affect the evolution of clustercentric radius-galaxy morphology/star formation activity (SFA) relation in clusters.

\begin{figure*}[tb]
    \centering
    \includegraphics[width=\textwidth]{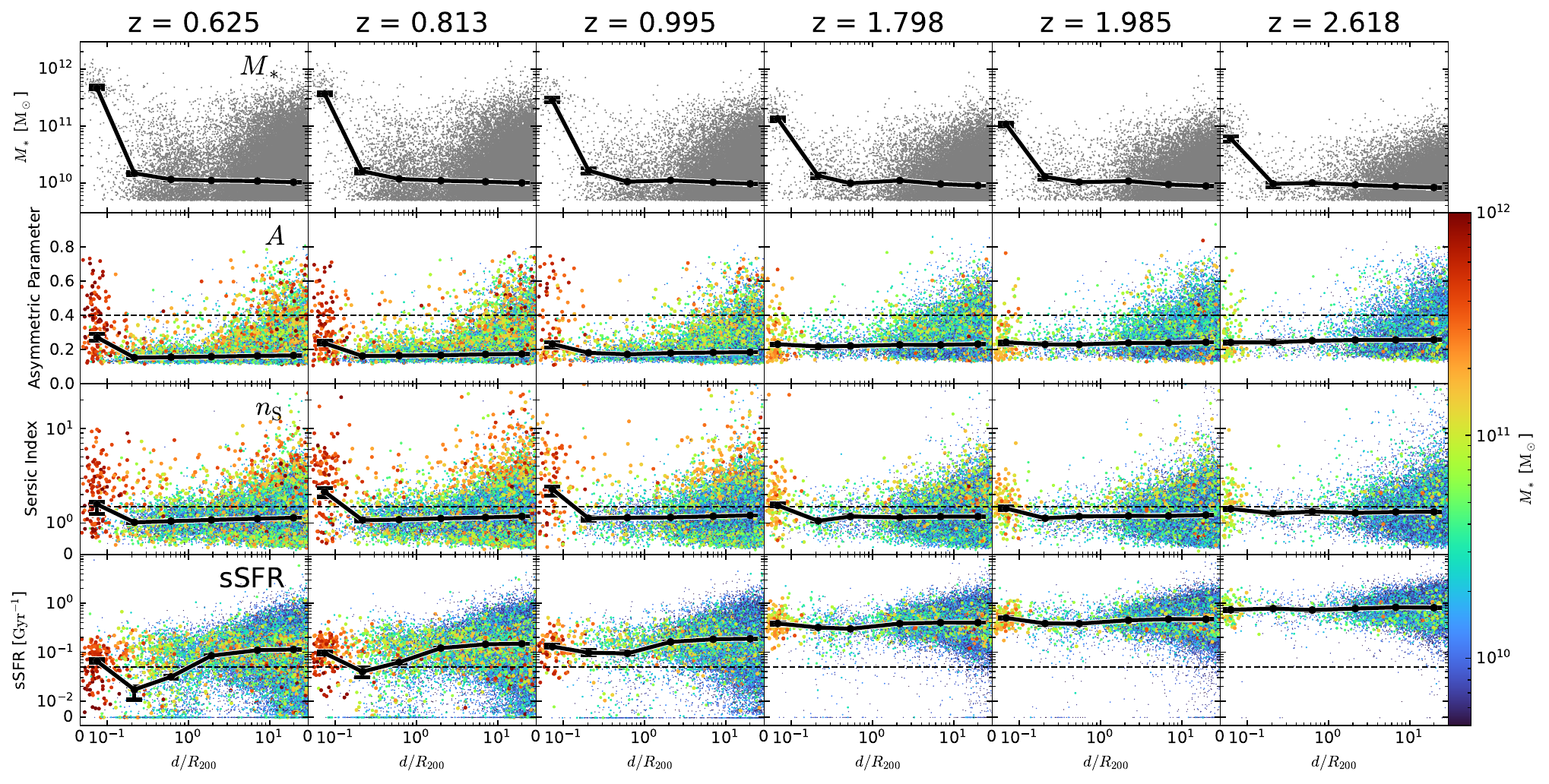}
    \caption{The clustercentric radial profiles of $\Mstar$, $\asym$, $\sersic$, and $\sSFR$ (from top to bottom) for HR5 galaxies with $\Mstar > 5 \times 10^9 \Msun$ (dots) at $z = 0.625$, $0.8$, $1$, $1.8$, $2$, and $2.6$ (from left to right). Colors: $\Mstar$ (not shown in the top panels). Error bars: the median and 68\% confidence interval derived from 9,999 bootstrapped samples from each $d/\Rvir$ bin.}
    \label{fig:radial_params_history}
\end{figure*}

Figure~\ref{fig:radial_params_history} shows the evolution of clustercentric radial profiles of stellar mass, asymmetry parameter, \Sersic index, and sSFR of HR5 galaxies with $\Mstar > 5 \times 10^9 \Msun$ from $z = 2.6$ to $\zf$. At $z=2.6$, the stellar mass of central galaxies is already nearly an order of magnitude larger than that of outskirts ones. On the other hand, there is only a slight indication of an excess in $\sersic$ at the center, and $\asym$ and sSFR do not show radial dependence. This again ensures that the stellar mass is not the only factor for the evolution of galaxy morphology.

At $z \lesssim 2$, all parameters $\asym$, $\sersic$, and sSFR gradually develop radial dependence. In particular, the majority of the central galaxies start to have the spheroidal morphology after $z\simeq 1.8$ (see the panel in the third row and the $z=1.798$ column of Figure~\ref{fig:radial_params_history}). Therefore, the morphology-clustercentric radius relation starts to emerge around $z \simeq 1.8$ when the galaxies populating the cluster central and outskirts regions begin to segregate into two classes: massive central spheroidal galaxies and low-mass outskirts disk galaxies.\footnote{Note that parameters including \Sersic index and sSFR from the HR5 snapshot right after $z = 1.5$ are not reliable due to the resolution refinement occurred at that time (see Appendix~\ref{sec:app_data_HR5}). This makes it hard for us to estimate the exact time when such two galaxy populations are fully developed.} At the final redshift of $z = 0.625$, the median $\sersic$ of galaxies at the cluster center has decreased. This happens because many relatively low-mass disk galaxies originated from the outskirts and far outside have now populated the central region of the cluster. Many of these disk galaxies are the quenched ones, as seen in the bottom left corner panel, and correspond to passive lenticular-type galaxies. It should be noted that lenticulars are as prevalent as ellipticals at the center of the massive clusters observed at low redshifts \citep[see Figures~4 and 5 of][]{dressler1980}.

The evolution of the radial profile of the asymmetry parameter also seems to be driven by the contrasting evolution of the central cluster galaxies versus the outskirts galaxies. The median value of $\asym$ for galaxies at the cluster center has shown a steady increase over time. This indicates that the cluster center is a region that promotes active galaxy-galaxy interactions. On the other hand, the median asymmetry of outskirts galaxies has slowly declined towards lower redshifts. Combining these two effects leads to a radial trend of the asymmetry parameter in clusters that becomes noticeable around $z \sim 1$ and continuously becomes stronger at lower redshifts.

However, the radial profile of sSFR evolves differently from the radial profile of other parameters. While $\Mstar$ and $\sersic$, dramatically evolve in the clustercentric radial profile during $1 \lesssim z \lesssim 2$, sSFR shows a nearly flat radial profile, mostly with $\sSFR > 0.05 \Gyr^{-1}$ (i.e., active galaxies). Even though the overall sSFR of all galaxies declines with cosmic time, the central galaxies ($d < 0.1 \Rvir$), dominated by massive galaxies, have managed to keep their sSFR in the active range until the final redshift. Although this may seem surprising, observationally, one can find many examples of BCGs at $0.625 \lesssim z \lesssim 1$ that are star-forming \citep[e.g.,][]{McDonald2016, Bonaventura2017}.

At the cluster outskirts, a non-negligible fraction of passive galaxies starts to appear in clusters around $z \simeq 1$. After $z\simeq 0.8$, a region where more than half of the galaxies are passive starts to appear and grow in size. We therefore consider $z\simeq 0.8$ as the epoch when the ARR emerges in our HR5 simulation. The exact epoch of the emergence of this relation must depend on the actual star formation history of the universe \citep[see, e.g.,][]{hwang2019}. As the decline of the sSFR in galaxies is determined by various astrophysical feedbacks, the epoch can vary among simulations adopting different feedback recipes and star formation history.

As discussed in Section~\ref{sec:results_global}, the inner-shell galaxies in the $0.1 < d/\Rvir < 0.3$ radial bin have experienced a faster decline in sSFR than central and outermost cluster galaxies. The infalling galaxies within the cluster undergo hydrodynamical interactions with the hot intracluster medium that may cut off the supply of cold gas required to sustain star formation \citep{Cowie1977, Bosch2008}. Additionally, hydrodynamic interactions with other cluster member galaxies may trigger starbursts and lead to the rapid consumption of cold gas in galaxies, eventually making them gas-poor and passive \citep{Moore1998, park2008, park2009}. On average, these galaxies should have had more time to interact with the cluster environment and the cumulative impact of intracluster medium, and their past multiple encounters with other cluster members may have resulted in a faster decline in sSFR than outer-shell galaxies \citep{hwang2018}. This difference is finally evident at $z\sim0.8$ where one can clearly see the emergence of the relation between sSFR and clustercentric distance, a monotonic increase of sSFR (bottom-left panel) with clustercentric distance for outskirts galaxies ($d > 0.1 \Rvir$). The central galaxies ($d < 0.1 \Rvir$), dominated by massive galaxies merging with infalling galaxies more frequently than low-mass ones, have managed to keep their sSFR in the active range until the final redshift.

\begin{figure}[tb]
    \centering
    \includegraphics[width=0.48\textwidth]{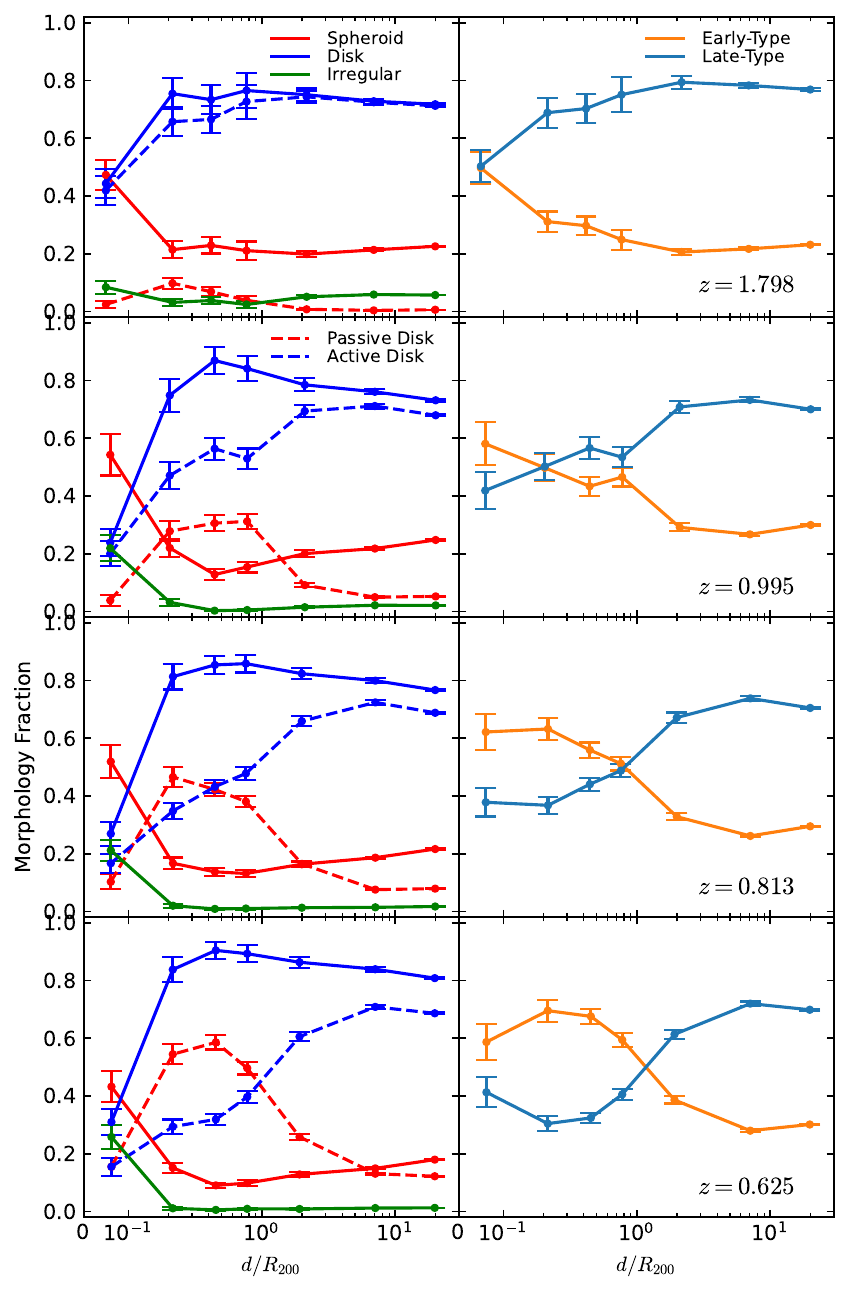}
    \vspace{-20pt}
    \caption{Evolution of clustercentric radial profiles of galaxy morphology (left) and type (right) fractions from $z = 1$ to $0.625$ (top to bottom). Error bars: Poisson error estimated from each radial bin.}
    \label{fig:morph_fraction_history}
\end{figure}

Figure~\ref{fig:morph_fraction_history} shows the evolution of clustercentric radial profiles of galaxy morphology (left column) and early/late type fractions from $z = 1$ (top row) to $0.625$. In all cases, disks constitute  $\sim 80\%$ of the entire galaxy population around cluster outskirts ($d > 0.3 \Rvir$), where spheroids are less than about $20\%$, and irregulars are nearly negligible. On the other hand, the fractions of both spheroids and irregulars increase rapidly around the cluster centers ($d < 0.1 \Rvir$).

As shown in Figure~\ref{fig:radial_params_history}, the radial distribution of early/late-type fraction evolves significantly along with the radial profile of sSFR between $z=1$ and $0.625$. First, the fraction of passive disks (red dashes in the left column) starts to increase significantly in the cluster outskirts after $z\simeq 1.8$. This phenomenon and the rapid increase of spheroidal fraction at the center result in the early/late type versus clustercentric radius relation emerging. The early-type dominance first appears at the center at $z\simeq 1.8$, and the region continuously expands to the outskirts all the way to $z=0.625$ (see the orange lines in the right column). Late-type galaxy fraction turns up slightly at the central region at $z=0.625$ in the bottom right panel because we assign all irregulars to the late type.

\section{Conclusions}\label{sec:conclusion}
We have studied the emergence and redshift evolution of the morphology-clustercentric distance (MRR), star formation activity-clustercentric distance (ARR), and early/late type-clustercentric distance relations using the \emph{Horizon Run 5} (HR5) cosmological hydrodynamical simulation. Our sample consists of 162 galaxy clusters with a total mass greater than $5 \times 10^{13} \Msun$ at $z = 0.625$ and all the galaxies with stellar mass $\Mstar > 5 \times 10^9 \Msun$ at various redshifts. We classify galaxies into morphological classes of spheroids, disks, and irregulars based on the \Sersic index and asymmetry parameter. We also divide galaxies into active and passive ones depending on their sSFR and into early and late types by considering both galaxy morphology and sSFR. Our main results are summarized as follows:
\begin{enumerate}
    \item The MRR starts emerging at $z \sim 1.8$ and already in place by $z \sim 1$. The main driver of this evolution seems to be the population of massive galaxies living near the cluster center ($d/\Rvir < 0.1$) that undergo significant morphological transitions from being a disk-dominated population at cosmic noon to becoming spheroidal galaxies by $z \sim 1.8$. This morphological transformation appears to be mainly driven by interactions and mergers with other galaxies. Galaxies residing in the cluster outskirts, however, do not undergo significant morphological changes, with a majority of them retaining their disk morphology till the final redshift of the HR5 simulation.
    \item The ARR emerges later than the MRR and is only apparent at $z \sim 0.8$ in the HR5 simulation. We find that passive disk galaxies residing in clusters are the main drivers of this relation. The median sSFR of cluster galaxies is similar to that of field galaxies at redshifts $z\gtrsim 1.8$.
    \item The combined effect of the above two relations drives the radial distribution of early and late-type galaxies in clusters. As the spheroidal morphological type starts to dominate the central region at $z\simeq 1.8$, the epoch of the emergence of the early/late type-clustercentric radius relation is also around $z=1.8$. However, the subsequent evolution of the two relations is quite different. Even though the region of spheroidal-type dominance remains confined at the center at all epochs, that of early-type dominance expands from the central region to the outskirts with time.
\end{enumerate}
The results presented in our work offer essential insights into galaxy evolution within clusters. While the properties of cluster galaxies and their scaling relations with cluster radius were known through observations, our simulation-based study reveals how and when these relations emerged. The deeper understanding gained from this work will aid in interpreting observational data in the future.

\begin{acknowledgments}
The authors thank an anonymous referee for helpful comments that improved the clarity of the paper.
S.E.H. is partly supported by the Korea Astronomy and Space Science Institute grant funded by the Korea government (MSIT) (No. 2024186901) and the grant funded by the Ministry of Science (No. 1711193044).
C.P. and J.K. are supported by KIAS Individual Grants (PG016903, KG039603) at Korea Institute for Advanced Study. C.P. is also supported by the National Research Foundation of Korea (NRF) grant funded by the Korea government (MSIT) (RS-2024-00360385).
P.K.M. is supported by KIAS Individual Grant (PG096701) at Korea Institute for Advanced Study.
Y.K. is supported by Korea Institute of Science and Technology Information (KISTI) under the institutional R\&D project (K24L2M1C4).
J.L. is supported by the National Research Foundation of Korea (NRF-2021R1C1C2011626).

This work was supported by the Supercomputing Center/Korea Institute of Science and Technology Information, with supercomputing resources including technical support (KSC-2013-G2-003), and the simulation data were transferred through a high-speed network provided by KREONET/GLORIAD. This work is also supported by the Center for Advanced Computation at Korea Institute for Advanced Study.

\end{acknowledgments}

%

\vspace{5mm}


\software{AstroPy \citep{astropy2013,astropy2018}, Matplotlib \citep{hunter2007}, Pandas \citep{mckinney2010, reback2020}}



\appendix

\section{Details on Horizon Run 5 Galaxy Data}\label{sec:app_data}

\subsection{Horizon Run 5 Simulation}\label{sec:app_data_HR5}

The linear power spectrum of HR5 \citep{lee2021} was made by the {\tt CAMB} package \citep{lewis2000} by adopting a flat $\Lambda$CDM cosmology in a concordance to {\it Planck} 2015 data \citep{planck2016} at $z = 200$. The corresponding initial condition was then produced by using the \texttt{MUSIC} package \citep{hahn2011} with second-order Lagrangian perturbation theory \citep[2LPT;][]{scoccimarro1998,lhuillier2014}. The initial condition of HR5 consists of two main parts: a low-resolution periodic cubic box of $(1048.6 \cMpc)^3$, and a high-resolution hydrodynamic region of $1048.6 \times 119.0 \times 127.2 \cMpc^3$ at the center of the background cubic box. Such an elongated shape of the high-resolution region was designed to see the impact of both large ($k \lesssim 10^{-2} \cMpc^{-1}$) and small scales ($k \sim 1 \pkpc^{-1})$ within a reasonable computation time. The background box was divided into $256^3$ coarse grids with a spatial resolution of $\Delta x = 4.10 \cMpc$, and only dark matter (DM) particles with a mass of $2.26 \times 10^{11} \Msun$ were used in the region. On the other hand, the high-resolution region was divided into $8192 \times 930 \times 994$ grids with $\Delta x = 128 \ckpc$, with an initial DM particle distribution with a mass of $6.89 \times 10^7 \Msun$ and the corresponding gas density in each cell. To compensate for the significant spatial resolution difference between the background box and the high-resolution region, HR5 added 4-level embedded padding regions in between, with spatial resolution increasing twice as the level increased by one.

The initial condition of HR5 was then evolved by a modified version of the adaptive mesh refinement code \texttt{RAMSES} \citep{teyssier2002}, with a heavy implementation of OpenMP parallelization on top of its original MPI. 
\texttt{RAMSES} solves the Poisson and Euler equations by using the Particle-Mesh method \citep{guillet2011} and the Harten-Lax-van Leer contact wave Riemann solver \citep{toro1994}, respectively. On the other hand, DM, stars, and massive black holes (MBHs) were assigned as separate particle species, while all three particle species were used to compute the mass density grids using a cloud-in-cell (CIC) scheme. 

During evolution, each cell was adaptively refined into eight higher-resolution cells when the mass within was less than eight times the DM particle mass at the given resolution. This resolution refinement was done until the spatial resolution reached $\Delta x \sim 1\pkpc$. Note that, to maintain the physical size of spatial resolution at the highest level throughout the evolution, HR5 added a new highest level of refinement at $z = 79$, $39$, $19$, $9$, $4$, and $1.5$ until the simulation reached its final redshift at $\zf = 0.625$.

Star particles were produced by a statistical star formation approach of \citet{rasera2006}, with a minimum star particle mass of $2.56 \times 10^6 \Msun$ at birth. Also, HR5 modeled the chemical evolution by adopting the model in \citet{few2012} and traced the abundance of hydrogen, oxygen, and iron based on the Chabrier initial mass function \citep{chabrier2003}. Several baryon-related properties, such as the star formation rate density, the galaxy stellar mass function, and the mass-metallicity relation, were calibrated by using test simulations (see Appendix~A of \citet{lee2021}).

During the evolution, the first 21 snapshots data were produced from $z = 200$ to $10$ with a uniform step of the logarithm of the expansion scale $\Delta \log a \approx 0.06$, where $a = (1+z)^{-1}$ is the expansion scale factor. Subsequently, 126 snapshots were produced from $z = 10$ to $\zf$ with a smaller uniform step of $\Delta \log a \approx 0.01$. As a result, 147 snapshot data were produced from HR5 from $z = 200$ to $0.625$.

\subsection{Halo \& Galaxy Identification}\label{sec:app_data_galaxy}

At each HR5 snapshot, halos were identified by applying the Friends-of-Friends (FoF) algorithm to the combination of DM particles, star particles, MBH particles, and gas cells. Here, each gas cell was considered as a single particle located at the center of the cell. Then, the linking length for each particle was assigned as a typical choice for finding virialized objects,
\begin{equation}
    \ell = 0.2 \left( \frac{m_{\rm p}}{\Om \rho_{\rm c}} \right)^{1/3} ~,
\end{equation}
where $m_{\rm p}$ is the particle mass, and $\rho_{\rm c}$ is the critical density at $z = 0$. A combined linking length $\ell_{12} = (\ell_1 + \ell_2)/2$ was used to link two particles with different masses or species.

After identifying halos, galaxies in HR5 were identified using a substructure finding algorithm \texttt{PGalF} \citep{lee2021}, an updated version of the \texttt{PSB} subhalo finder \citep{kim2006} originally developed for $N$-body simulations \citep{kim2009, kim2015}. \texttt{PGalF} identifies galaxies as groups of star particles embedded in halos as follows. First, the local stellar mass density is calculated for each star particle by applying the $W_4$ smoothed particle hydrodynamics (SPH) density kernel to all star particles within the same halo. Then, robust local maxima of local stellar mass density are assigned as the galaxy center. Here, we assume that galaxies should contain at least ten star particles, which makes the minimum stellar mass of galaxies $\sim 2.6 \times 10^7 \Msun$. After that, nearby star particles are assigned to the galaxy by applying the density cut from the watershed algorithm. Finally, the membership star particle candidates outside tidal boundaries or having positive total energy are rejected from each galaxy. Readers may refer to \citet{lee2021} and \cite{park2022} for details.

\subsection{Merger Tree}\label{sec:app_data_tree}

To understand the history of halos and galaxies in HR5, their merger trees were constructed using the 147 snapshots of HR5 with different techniques.
First, halos merger trees were constructed using the most bound particles (MBPs), a DM particle with the minimum total energy within the given halo \citep{hong2016}. If a halo of a certain snapshot contains the MBP of another halo that existed at the previous snapshot, then the former halo is the ``descendant'' of the latter one, while the latter is a ``progenitor'' of the former. When two or more progenitors exist for a single halo (e.g., due to the halo merger event), the progenitor having the greatest total mass is regarded as the main progenitor, and we set the branch that consists of the main progenitor and its descendant as the main branch.

While the same technique using MBPs can be used to construct the merger trees of galaxies, \citet{lee2021} and \citet{park2022} found that galaxy merger trees constructed by only MBPs do not show the proper inheritance of star particles in numerous cases. Instead, we apply an updated version of another tree-building code \texttt{ySAMtm} \citep{jung2014, lee2014, park2022} to star particles to construct the merger trees of galaxies as follows. First, \texttt{ySAMtm} maps galaxies from one snapshot to those in the next snapshot that share at least a single star particle. Then, the galaxy at the previous/next snapshot that shares the largest number of star particles with a galaxy becomes the main progenitor/descendant. If a galaxy is the main descendant of another galaxy in the previous snapshot and the latter is the main progenitor of the former, we set the branch connecting these two galaxies as the main branch.

Note that, however, the actual merger trees of halos and galaxies in HR5 have more complicated situations than described above in numerous cases. For example, the descendant of a certain halo/galaxy may appear several steps after instead of the very next step. Or, fly-by events that two or more progenitors merge toward a single halo/galaxy and then split into multiple descendants may happen. The reader may refer to \citet{park2022} for details of how our merger tree methods work in such cases.

\subsection{Morphology Classification}\label{sec:app_data_morph}

We measure the ``morphology'' of HR5 galaxies by using the asymmetry and the radial profile of their stellar mass density distribution by adopting the same method described in \citet{park2022}. First, the asymmetric parameter $\asym$ of a galaxy is defined in a similar way to \citet{conselice2000}, 
\begin{equation}
   \asym = \frac{\sum_{{\bf x}} w({\bf x}) | \rho_\ast ({\bf x}) - \rho_\ast ({\bf \bar{x}}) |}{\sum_{{\bf x}} w({\bf x}) \rho_\ast ({\bf x}) } \, ,
\end{equation}
where ${\bf x}$ is the positions of grid cells around the galaxy, and ${\bf \bar{x}} \equiv 2{\bf x}_{\rm cen} - {\bf x}$ is the opposite position in terms of the galaxy center (e.g., local density peak). $\rho_\ast ({\bf x})$ is the stellar mass density at ${\bf x}$, calculated by applying a spline kernel with 100 nearby star particles. Also, $w({\bf x})$ is the weight for assigning the member grid cells of the galaxy, which is one if either ${\bf x}$ or ${\bf \bar{x}}$ contains more than ten star particle and zero otherwise. We also set $w$ in the central region with a diameter of $1.6\pkpc$ as zero, as this region is not well resolved in HR5. We calculate $\asym$ for galaxies with stellar mass greater than $2 \times 10^9 \Msun$ or more than $\sim 1000$ star particles. \citet{park2022} calculated the asymmetric parameter of galaxies at $z = 5 - 7$ in HR5 and found that its distribution is well separated around $\asym \approx 0.4 - 0.5$. Therefore, same as the reference, we classify galaxies with $\asym > 0.4$ as asymmetric galaxies (or ``irregulars''), while those with $\asym < 0.4$ are called symmetric galaxies.

For symmetric galaxies, the \Sersic index $\sersic$ is defined as \citep{sersic1963}
\begin{equation}
    \Sigma(R) = \Sigma_0 \exp \left[-kR^{1/\sersic} \right] ~.
\end{equation}
Here, $\Sigma(R)$ is the radial profile of the stellar mass density projected on the galactic plane defined by the major and intermediate axes, $R$ is the radius in the galactic plane, and $k$ is a parameter related to the overall slope of the radial profile. In practice, we find the major and intermediate axes of symmetric galaxies by diagonalizing the moment of inertia tensor of the member star particles. Then, we fit the radial density profile over the radial interval from $0.8\pkpc$ to $R_{90}$, a radius where 90\% of total star particles are contained, to find the best-fit value of the \Sersic index. \citet{park2022} found that the galaxies at the cosmic morning of HR5 are well separated into two populations around $\sersic \approx 1.5$. Therefore, we classify symmetric galaxies with $\sersic < 1.5$ as disks, while symmetric galaxies with $\sersic > 1.5$ are called spheroids. Note that some symmetric galaxies cannot have the best-fit $\sersic$ because their radial profile is inconsistent with the \Sersic model, and we call such galaxies irregulars too.

\subsection{Active/Passive Classification}
\label{sec:app_data_active_passive}

\begin{figure*}[tb]
    \centering
    \includegraphics[width=0.9\textwidth]{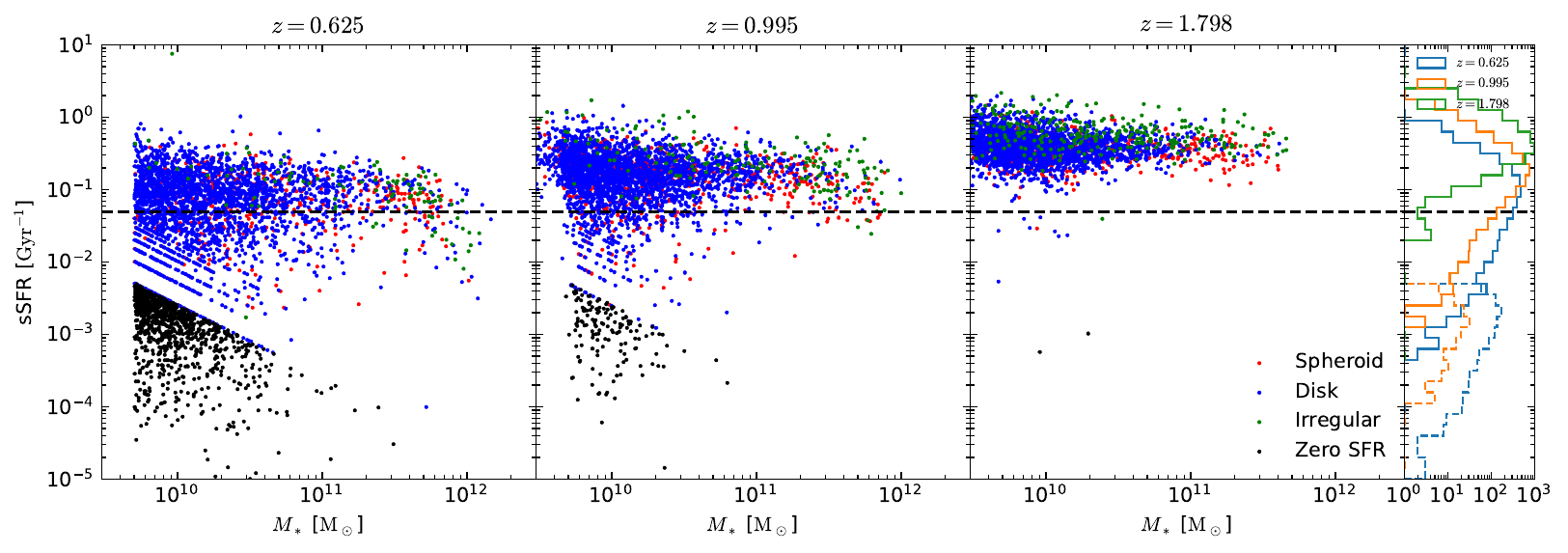}
    \caption{Evolution of stellar mass vs. specific star formation rate (sSFR) relation of HR5 cluster galaxies at $z = 0.625$, $1$, and $1.8$ (from left to right). Dots: cluster galaxies with color showing their morphology at a given time. Colored steps at the rightmost panel: histograms of the sSFR of cluster galaxies at different epochs. Black dots and dashed steps: cluster galaxies whose star formation rate (SFR) is marked as zero because of the resolution limit of star particles. We apply uniformly random values of SFR between 0 and the minimum nonzero value from the simulation only for this plot. Dashed line: $\sSFR = 0.05 \Gyr^{-1}$, which we use to distinguish active and passive galaxies in this paper.}
    \label{fig:mstar_ssfr}
\end{figure*}

We divide our galaxy sample into ``active'' and ``passive'' galaxies, depending on their specific star formation rate ($\sSFR \equiv {\rm SFR} / \Mstar$). Figure~\ref{fig:mstar_ssfr} shows the relationship between our cluster galaxies' stellar mass and sSFR at redshifts $z = 0.625$, $1$, and $1.8$. Some previous works have suggested the classification of passive and active galaxies using a time-dependent sSFR threshold that separates the dichotomy sSFR distribution \citep[e.g.,][]{katsianis2021}. However, we could not find such dichotomy distribution from HR5 cluster galaxies, possibly due to the limitation of the time and spatial resolution to see the lower tail of the sSFR distribution (see the black dots in Figure~\ref{fig:mstar_ssfr} where SFR is assigned to zero). Instead, we compare the $\Mstar$-$\sSFR$ distribution in different epochs and set a global sSFR threshold so that most galaxies are active at high-$z$ and numerous low-mass galaxies at low-$z$ are passive. We found that $\sSFR \approx 0.05 \Gyr^{-1}$ satisfies the above conditions and define active and passive galaxies in HR5 as those with $\sSFR > 0.05 \Gyr^{-1}$ and $\sSFR < 0.05 \Gyr^{-1}$, respectively (dashed line). Note that our choice of sSFR threshold could be different from conventional choices from observations \citep[e.g., $\sSFR \approx 0.01 \Gyr^{-1}$ in][]{cassata2010, ilbert2010, tamburri2014}, mainly because our sSFR threshold heavily depends on the star formation activity-related parameter calibration of HR5.

\section{Evolutions of Parameter-Parameter Relations for Selected Galaxies}
\label{sec:app_flow_precise}

\begin{figure*}[tb]
    \centering
    \includegraphics[height=0.42\textwidth]{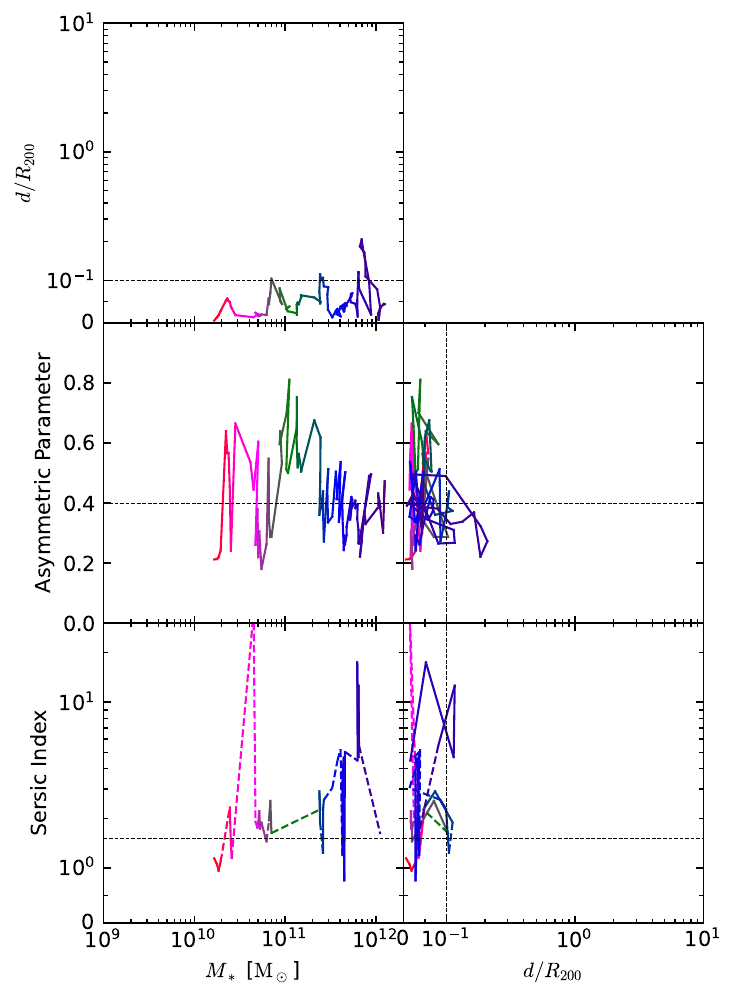}
    \includegraphics[height=0.42\textwidth]{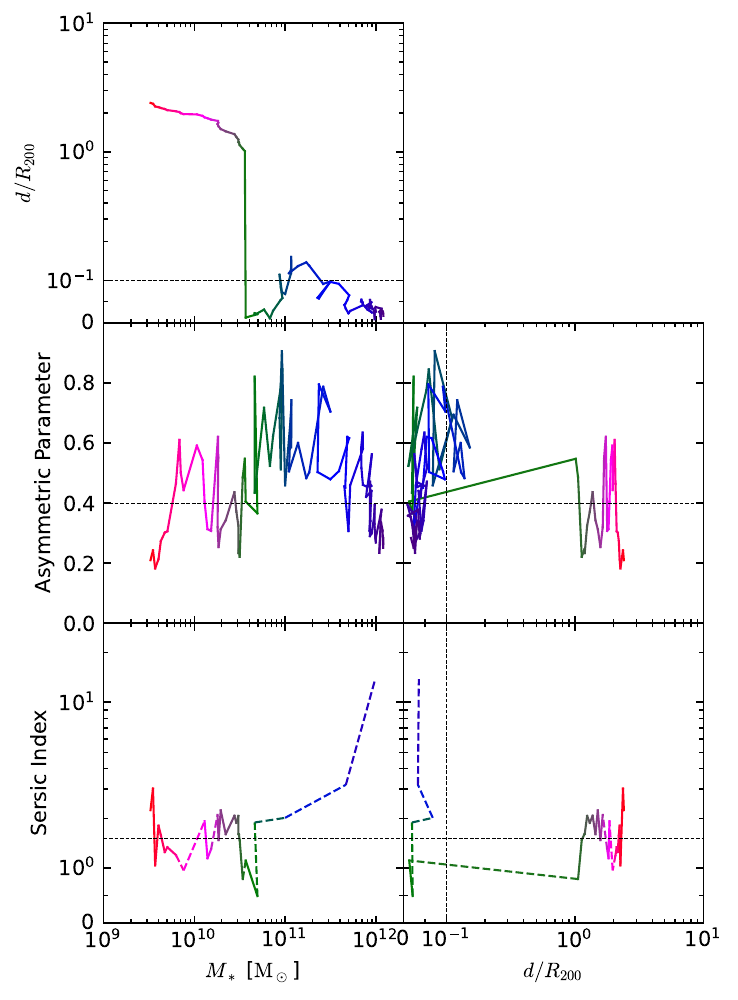}
    \includegraphics[height=0.42\textwidth]{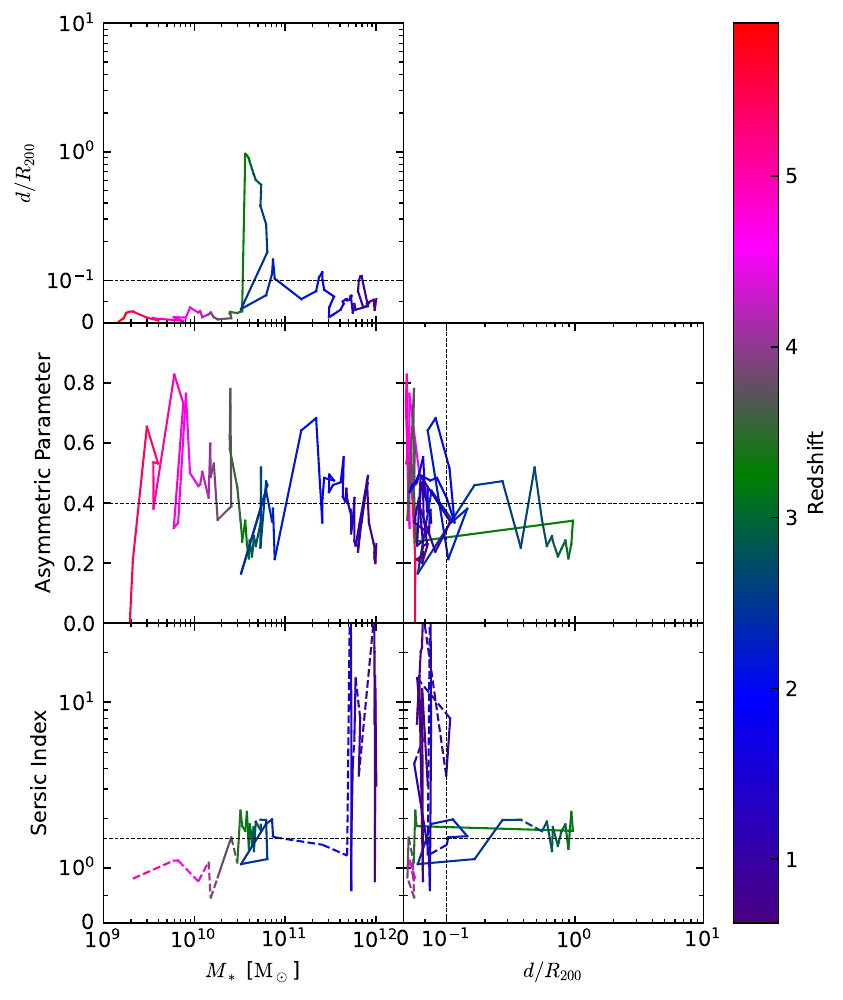}
    \caption{Similar to Figure~\ref{fig:flow}, showing the detailed evolution of various parameter-parameter relations for the 1st (left), 2nd (middle), and fifth (right) most massive HR5 cluster galaxies from $z = 6$ (red) to $0.625$ (purple). The \Sersic parameter with null values at snapshots with $\asym > 0.4$ is shown as dashed lines.}
    \label{fig:flow_precise_massive}
\end{figure*}

\begin{figure*}[tb]
    \centering
    \includegraphics[height=0.42\textwidth]{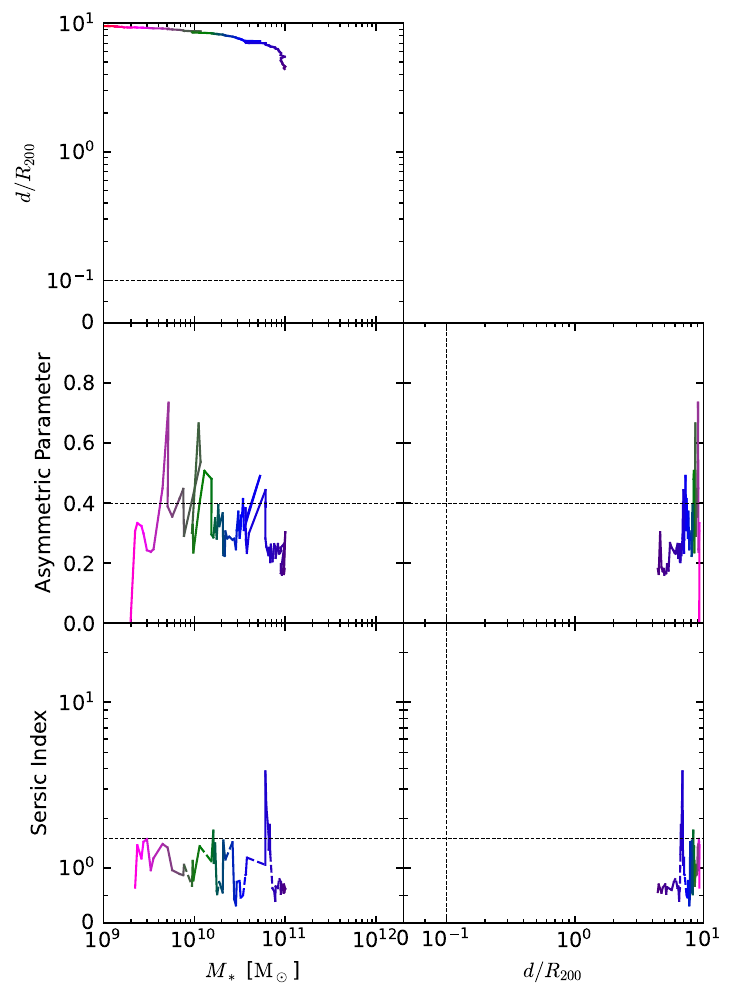}
    \includegraphics[height=0.42\textwidth]{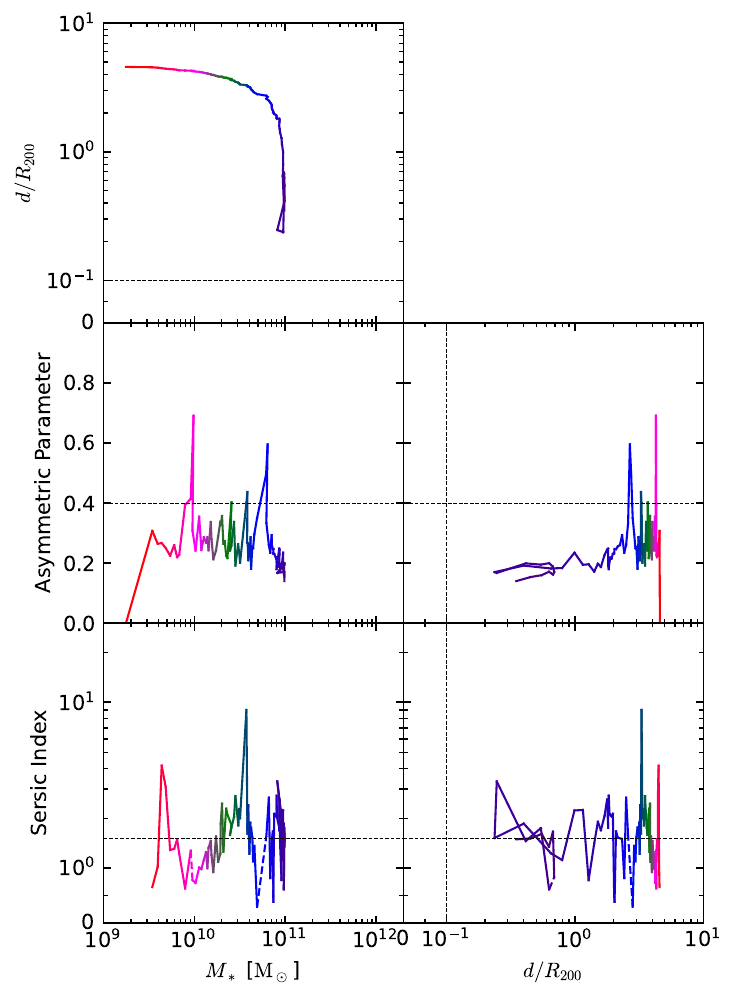}
    \includegraphics[height=0.42\textwidth]{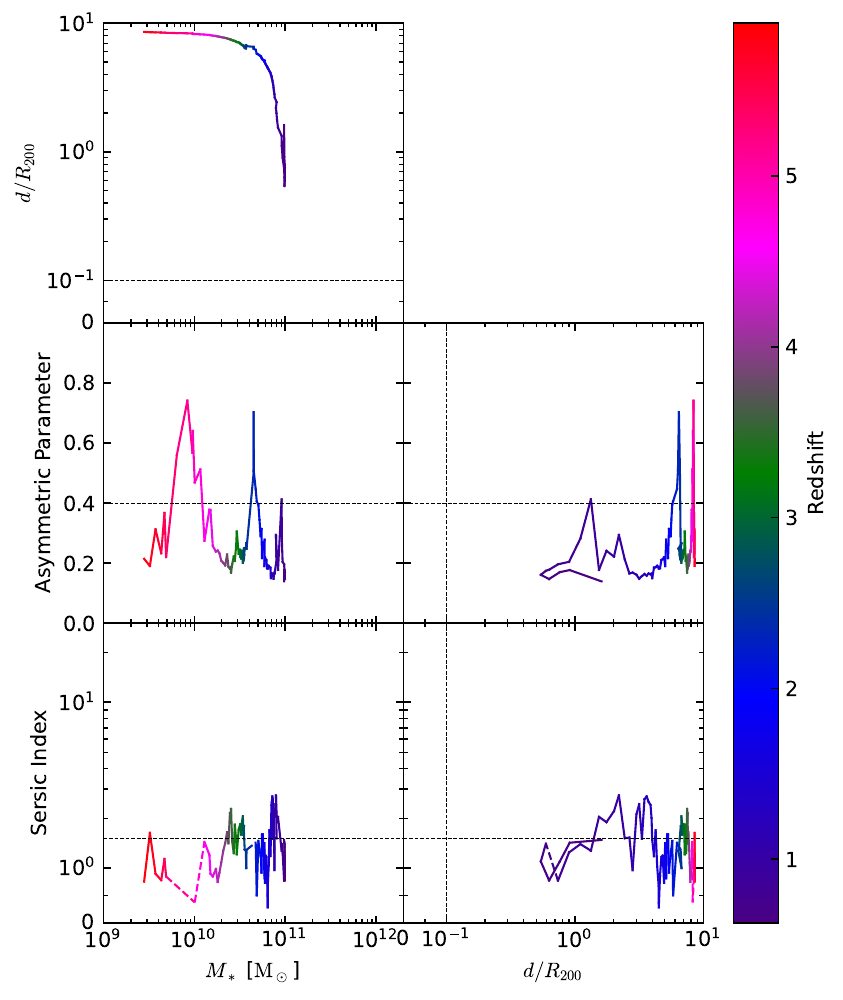}
    \caption{Similar to Figure~\ref{fig:flow_precise_massive}, showing the three selected HR5 cluster galaxies with $\Mstar < 10^{11} \Msun$ and $d/\Rvir > 0.3$ at $z = 0.625$.}
    \label{fig:flow_precise_lowmass_outskirt}
\end{figure*}

Figures~\ref{fig:flow_precise_massive} and \ref{fig:flow_precise_lowmass_outskirt} show the detailed evolution of parameter-parameter relations among $\Mstar$, $d/\Rvir$, $\asym$, and $\sersic$ of three very massive HR5 cluster galaxies and low-mass outskirts HR5 cluster galaxies, respectively, from $z = 6$ to $0.625$.

Most of the very massive HR5 cluster galaxies have reached near the cluster centers ($d < 0.1 \Rvir$) before or around cosmic noon ($z \gtrsim 3$) and have been there since then. Their normalized clustercentric radii oscillate over time (or $\Mstar$; see top-left panels of Figure~\ref{fig:flow_precise_massive}), possibly related to their orbital motions passing through the cluster centers. Such oscillations at $\Mstar$-$(d/\Rvir)$ are correlated to oscillations at $\Mstar$-$\asym$, especially for those that reach peaks at $\asym > 0.4$, indicating major merger events or tidal disruptions (middle-left panels). Not each major merger or tidal disruption event guarantees the morphology change of very massive galaxies from disks to spheroids, but multiple passes of such events around cluster centers would stochastically change their morphology (see bottom-left).

On the other hand, low-mass outskirts HR5 cluster galaxies ($\Mstar < 10^{11} \Msun$; $d/\Rvir > 0.3$ at $z = \zf$) have not experienced such dramatic morphology change over time. Note that they also have had several events of $\asym > 0.4$ with a comparable number to the very massive HR5 cluster galaxies, but most of such events occurred before the galaxies entered the clusters ($d > \Rvir$).

\section{Clustercentric radius vs. local background density on the galaxy morphology \& star formation activity}

Numerous literatures have studied the relation between the local background density, instead of clustercentric radius, and the galaxy morphology/star formation activity (SFA) \citep[see, e.g.,][]{dressler1980, dressler1997, treu2003, goto2003, postman2005, gavazzi2010}. However, \citet{park2009} have found that at fixed mass or absolute magnitude, it is actually clustercentric radius $d$ and the nearest-neighbor environment that governs the properties of cluster galaxies and that the dependence on the local background density appears to exist just because of its statistical correlation with $d$ and nearest-neighbor environment. Here, the nearest-neighbor environment means the distance and morphology of the nearest ``neighbor'' galaxy \citep{Park&Choi2009, park2009}.

\begin{figure*}[tb]
    \centering
    \includegraphics[width=0.8\textwidth]{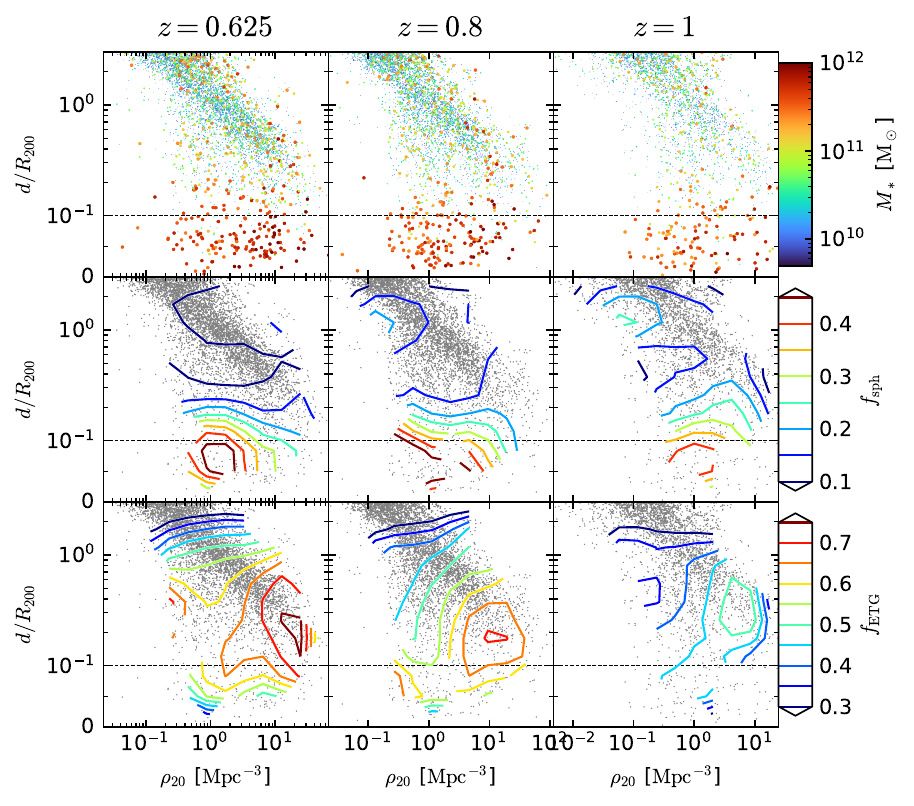}
    \vspace{-10pt}
    \caption{Relation between the local galaxy number density measured by 20 nearest neighbors ($\rho_{20}$) and the normalized clustercentric radius for HR5 galaxies at $z = 0.625$, $0.8$, and $1$ (from left to right). Each dot corresponds to a single HR5 galaxy with $\Mstar > 5 \times 10^9 \Msun$ at a given redshift. Top: The size and color indicate the stellar mass. Middle: Contours of the fraction of spheroid galaxies ($f_{\rm sph}$) are overlayed. Bottom: Same as middle, except for the fraction of early-type galaxies ($f_{\rm ETG}$).}
    \label{fig:dens_rad_Mstar_morph}
\end{figure*}

Figure~\ref{fig:dens_rad_Mstar_morph} shows the relation between the local galaxy number density ($\rho_{20}$) and $d/\Rvir$ of HR5 galaxies with $\Mstar > 5 \times 10^9 \Msun$ at $z = 0.625$, $0.8$, and $1$. Here, since we want to check the profile of galaxy morphology and SFA type at each snapshot, we use ``all'' galaxies found in each HR5 snapshot data above a stellar mass threshold rather than the ``cluster galaxies'' defined as the main progenitors of the galaxies found at $z = \zf$. (There are galaxies that existed near the clusters at high redshifts but did not survive to become a cluster galaxy at $z = \zf$.) Also, the normalized clustercentric radius of a given HR5 galaxy is assigned by finding the corresponding HR5 cluster that has the smallest value of $d/\Rvir$. Finally, we define the local galaxy number density of a given HR5 galaxy by applying the spline function kernel to the 20 nearest HR5 galaxies, including the galaxy itself \citep{monaghan1985,monaghan1992,park2007,song2016}:
\begin{equation}
   \rho_{20} \equiv \sum_{i=1}^{20} W(r_i,h_{\rm spl}) \, ,
\end{equation}
where $r_i$ is the distance between a given HR5 galaxy and $i$-th nearest neighbor, $h_{\rm spl} \equiv r_{20} / 2$ is the smoothing length, and
\begin{equation}
   W(r_i,h_{\rm spl}) \equiv \frac{1}{\pi h_{\rm spl}^3} \left\{  
   \begin{array}{ll}
    1 - \frac{3}{2} q_i^2 + \frac{3}{4} q_i^3   & \textrm{ if } 0 \leq q_i < 1 \\
    \frac{1}{4} (2 - q_i )^3    &  \textrm{ if } 1 \leq q_i \leq 2 \\
    0 & \textrm{ otherwise}
   \end{array} \right.
\end{equation}
is the spline function kernel with $q_i \equiv r_i / h_{\rm spl}$.\footnote{Note that most observational studies using the local density use the 2D number density ($\Sigma$), mainly due to the observational difficulties of finding positions in the line-of-sight direction.}

As expected, there exists a strong correlation between $\rho_{20}$ and $d/\Rvir$ for galaxies in and around clusters, as demonstrated by the narrow band of the galaxy distribution. If we only focus on the galaxies within clusters ($d < \Rvir$), the fraction of spheroids ($f_{\rm sph}$) depends mainly on $d /\Rvir$, which is evidenced by the roughly horizontal contour lines (middle panels). On the other hand, $f_{\rm ETG}$ appears to depend largely on $\rho_{20}$, which is indicated by roughly vertical contours at $0.1 \Rvir < d < \Rvir$ (bottom panels). According to \citet{park2009}, it is actually the hydrodynamic interaction with the nearest neighbor galaxy rather than the background density that affects the star formation properties of cluster galaxies \citep[see Figure~10 of][]{park2009}. The statistical correlation of background density with distance and morphology of the nearest neighbor produces the apparent relation between $\rho_{20}$ and $f_{\rm ETG}$.

\section{Linking between clustercentric morphology and SFR profiles}

The specific star formation rate profile is an impact of the clustercentric radius ($d$) on the galaxies' star formation rate, essentially a measure of the instantaneous availability of cold gas within the galaxy. Conversely, morphology is the integrated impact of stochastic perturbation on the 
stellar structure of the galaxy \citep{park2009}. As such, the two scaling laws are quite distinct a priori (driven by different processes operating on different timescales) and need not appear at the same cosmic time or scale similarly with a clustercentric radius. The former is impacted by (1) close encounters with cluster early-type galaxies, which can deplete cold gas in the late-type host galaxy   \citep{park2009}, (2) ram pressure stripping with an efficiency that depends on the hot gas density within the cluster; (3) disconnection from the cold web. The latter is  impacted by cooling processes (new stars formed on quasicircular orbits within the galaxy), as well as internal and external stochastic heating processes. These involve supernova explosions, turbulence, ram pressure heating (which displaces the gas within the galaxy), and potential fluctuations reflecting the impact of tides (which is clearly $d$-dependent). We can anticipate qualitatively that tidally driven gravitational heating leads to a clustercentric gradient of morphology.  Similarly, star formation quenching should increase toward small $d$ because encounters with early-type galaxies become more frequent and stronger and cluster ram pressure increases.

Let us be a bit more specific and assume that the impact of sinking in the cluster is sufficiently temperate so that it can be treated quasi-linearly, hence within the realm of kinetic theory \citep{1988MNRAS.230..597B,2006MNRAS.368.1657P,10.1051/0004-6361/201527052}. Let us also assume that we can stack galaxies of fixed mass, cosmic age, and clustercentric distances to  build a model for the cosmic evolution of the morphology of a typical galaxy.

For a given cosmic time $t$, let us assume that the mean field of the galaxy ($\psi$) is known so that the corresponding angle-actions $\mathbf{\varphi}$ and $\mathbf{J}$ are known. This mapping may need to be adapted adiabatically across cosmic time. The ensemble-averaged  stellar distribution of stars within the galaxy situated at clustercentric radius $d$, $f(t;d)$, satisfies a sourced Fokker Planck (sFP) equation in action space \citep{2006MNRAS.368.1657P}
\begin{equation}
\frac{\partial f(t;d)}{\partial t}= \frac{\partial}{\partial \mathbf{J}}\cdot \mathbf{D}(d) \cdot \frac{\partial f(t;d)}{\partial \mathbf{J}} +s(\mathbf{J};d) \,,\label{eq:FP}
\end{equation}
where $s(\mathbf{J};d)$ is the ensemble-averaged source term reflecting the production of stars, while $\mathbf{D}$ is the diffusion tensor
\begin{equation}
\mathbf{D}(d) = \sum_\mathbf{k}\, \mathbf{k}\otimes \mathbf{k}\langle \left| \delta \hat \psi_\mathbf{k}(\mathbf{J},\omega= \mathbf{\Omega}\cdot \mathbf{k})  \right|^2 \rangle_d \,,
\end{equation}
where $\delta \hat \psi_\mathbf{k}(\omega)$ is the Fourier transform with regard to the angles and time of all internal and external potential fluctuations along the orbit, $\mathbf{\Omega}$ is the frequencies of the stars, and $\mathbf{k}$ is the integer harmonic numbers conjugate to the angles. Eq.~(\ref{eq:FP}) can be written as an integral equation 
\begin{equation}
f(\mathbf{J},t; d) =\int \mathrm{d} \mathbf{J}'  \mathrm{d}t' \, G(\mathbf{J},t|\mathbf{J}',t'; d) s(\mathbf{J}',t'; d) \,, \label{eq:inteq}
\end{equation}
via the corresponding Green function, $G(\mathbf{J},t|\mathbf{J}',t'; d) $.

In the phase-averaged Chandrasekar approximation \citep{2024arXiv240201506T}, the diffusion coefficient tensor can be parametrized via some simple local  power spectrum $P(\omega,r=|\mathbf{r}|; k,d)$ of potential fluctuations within the disk at position $\mathbf{r}(\varphi)$, belonging to a galaxy at a clustercentric radius $d$:
\begin{equation}
\mathbf{D}(d) \approx \int\! \mathrm{d}\varphi\mathrm{d} \mathbf{k} \, \mathbf{k} \otimes\mathbf{k} P(\omega=\mathbf{k}\cdot \mathbf{v},r; k,d)\,, \label{eq:localD}
\end{equation}
where $\varphi$ and $\mathbf{v}(\varphi)$ are the orbital phase and the velocity of the star along its orbit $\mathbf{J}$, respectively. Eq.~(\ref{eq:localD}) captures that only galactic stars that resonate ($\omega=\mathbf{k}\cdot \mathbf{v}$) with the noise structure will be significantly impacted.

Given $s(\mathbf{J},t; d)$ and $\mathbf{D}(d)$, the integral equation Eq.~(\ref{eq:inteq}) gives access to the ensemble-averaged distribution function of the galaxy as a function of $d$ and cosmic time ($t$).
The surface density, $\Sigma(r,t; d)$ of the average galaxy can be deduced by simple integration,  
\begin{equation}
\Sigma(r,t; d)=\int \mathrm{d}^3 v \, \mathrm{d} z f(\mathbf{J},t; d) \,,
\end{equation}
while its radially averaged \Sersic index is
\begin{equation}
\sersic(d,t) \equiv -\langle \mathrm{d} \log \Sigma(r,t;d) /\mathrm{d} \log r \rangle_r^{-1} ~.
\label{eq:def-sersic}
\end{equation}
Therefore, the fraction of spheroids should vary with cosmic time and clustercentric radius. It would be affected by both heating, driven by tides, and other stochastic processes in the galaxy, which broaden the disk and disk rejuvenation. This can be expressed as
\begin{equation}
s(\mathbf{J},t;d)\propto \Sigma(r,t;d)\, \sSFR(t;d)\delta_\mathrm{D}(J_z)\delta_\mathrm{D}(J_r)\,, \label{eq:defSR}
\end{equation} where $J_z$ and $J_r$ are the vertical and radial actions, and 
$\sSFR(t;d)$ is typically stronger at larger $d$ and earlier time. The ARR discussed in the main text is a direct measure of $\partial \sSFR / \partial d >0$, while the MRR is a measure of $\partial \sersic / \partial d <0$. Eqs.~(\ref{eq:inteq})--(\ref{eq:defSR}) quantify how the two are related.
 
In practice, baryonic processes also couple to the dynamics by boosting  internally amplified tides within the disk, which is partially self-gravitating. As such, one should, at the next order, take into account the fact that the power spectrum of fluctuation is dressed by the response of the cold disk \citep{2017MNRAS.471.2642F}, which in turn stiffens the disk \citep{2021ApJS..254...27K} and accelerates star formation \citep{2019ApJ...883...25P}. This maintains the disk thin and close to marginal stability, as long as cold gas is available to maintain self-regulation \citep{Pichon2023-bs, Pichon2023-dk}. When the galaxy sinks within the cluster to radii where cold gas may not be available in sufficient amounts (e.g., because of close encounters with early type galaxies and ram pressure stripping), self-regulation should stop, and star-forming activity type transformation (from late type to lenticular) should be unavoidable.
 
Ultimately, the measured scales and timescales at which the ARR and MRR are established will, therefore, depend on the baryonic accretion history of the cluster. Therefore, it is not surprising that the latter is related to cosmic noon \citep{2023ApJ...953..119P}, while the former is a fraction of $\Rvir$  corresponding to the typical size of hot bubbles of gas in the intracluster medium. To a lesser extent, e.g., the stellar and AGN feedback models or the star formation recipes would impact heating \citep{Jackson}. Lack of resolution will also truncate the noise power spectra, artificially stabilizing the disks.


\bibliography{bib}{}
\bibliographystyle{aasjournal}



\end{document}